\documentclass[11pt,a4paper]{article}
\pdfoutput=1
\usepackage{jheppub}

\usepackage{multirow, graphicx,amssymb,url,mathrsfs,amsmath}
\usepackage{wrapfig,boxedminipage,setspace,subfigure,epsfig}
\usepackage{amsxtra,amstext,latexsym,dsfont,amsfonts}
\usepackage{color,eucal}
\usepackage[dvipsnames]{xcolor}
\usepackage{float}
\usepackage{slashed,comment}
\usepackage{tabularx, array}

\newcolumntype{L}[1]{>{\raggedright\arraybackslash}p{#1}}
\newcolumntype{C}[1]{>{\centering\arraybackslash}p{#1}}
\newcolumntype{R}[1]{>{\raggedleft\arraybackslash}p{#1}}

\newcommand{\dd}{\mathrm{d}}

\title{Upper bound of the charge diffusion constant in holography}

\author[a]{Kyoung-Bum Huh,}
\author[b,c]{Hyun-Sik Jeong,}
\author[a]{Keun-Young Kim,}
\author[b,c]{and Ya-Wen Sun}

\emailAdd{hkabell1689@gm.gist.ac.kr}
\emailAdd{hyunsik@ucas.ac.cn}
\emailAdd{fortoe@gist.ac.kr}
\emailAdd{yawen.sun@ucas.ac.cn}

\affiliation[a]{School of Physics and Chemistry, Gwangju Institute of Science and Technology, \\
123 Cheomdan-gwagiro, Gwangju 61005, Korea}
\affiliation[b]{School of physics $\&$ CAS Center for Excellence in Topological Quantum Computation, University of Chinese Academy of Sciences, Zhongguancun east road 80, Beijing 100049, China}
\affiliation[c]{Kavli Institute for Theoretical Sciences, University of Chinese Academy of Sciences, \\ Zhongguancun east road 80, Beijing 100049, China}

\abstract{
We investigate the upper bound of charge diffusion constant in holography. For this purpose, we apply the conjectured upper bound proposal related to the equilibration scales ($\omega_{\text{eq}}, k_{\text{eq}}$) to the Einstein-Maxwell-Axion model. ($\omega_{\text{eq}}, k_{\text{eq}}$) is defined as the collision point between the diffusive hydrodynamic mode and the first non-hydrodynamic mode, {giving rise to} the upper bound of the diffusion constant $D$ at low temperature $T$ as $D = \omega_{\text{eq}}/k_{\text{eq}}^2$.
We show that the upper bound proposal also works for the charge diffusion and ($\omega_{\text{eq}}, k_{\text{eq}}$), at low $T$, is determined by $D$ and the scaling dimension $\Delta(0)$ of an infra-red operator as $(\omega_{\text{eq}}, \, k_{\text{eq}}^2) \,=\, (2 \pi T \Delta(0) \,, \omega_{\text{eq}}/D)$, as for other diffusion constants.
However, for the charge diffusion, we find that the collision occurs at real $k_{\text{eq}}$, while it is complex for other diffusions.
In order to examine the universality of the conjectured upper bound, we also introduce a higher derivative coupling to the Einstein-Maxwell-Axion model. This coupling is particularly interesting since it leads to the violation of the \textit{lower} bound of the charge diffusion constant so the correction may also have effects on the \textit{upper} bound of the charge diffusion. We find that the higher derivative coupling does not affect the upper bound so that the conjectured upper bound would not be easily violated.
}

\begin{document}
\maketitle

\section{Introduction}

Holographic methods (gauge/gravity duality)~\cite{Hartnoll:2016apf,Zaanen:2015oix,Ammon:2015wua,Baggioli:2019rrs} have been providing new and effective ways to investigate the universal features in the transport properties of the strongly correlated systems. For example, in strongly coupled systems like strange metals, the resistivity ($\rho$) is universally linear in temperature ($T$), $\rho \sim T$, which is in contrary to $\rho \sim T^2$ from the Fermi liquid theory. The linear-$T$-resistivity ($\rho \sim T$) has been explored in holography. See \cite{Hartnoll:2016apf} and references therein for the achievements with methodologies.
Another example is Homes' law. Strange metals would undergo a phase transition to the high temperature superconductor with a universal relation between the superfluid density ($\rho_s$), the critical temperature ($T_c$), and the DC electric conductivity ($\sigma_{DC}$), i.e. the so called Homes' law: $C := \rho_s(T=0)/(\sigma_{DC}(T_c)T_c)$, where $C$ is universal and independent of the components of superconducting materials. For holographic studies of Homes' law, see \cite{Erdmenger:2015qqa,Kim:2015dna,Kim:2016hzi,Kim:2016jjk}.

In this paper, we focus on another universal property of strongly coupled systems in holography: the identification of a universal bound of the diffusion constant ($D$).

One of the famous examples of the holographic bound might be the Kovtun-Son-Starinets (KSS) bound~\cite{Policastro:2001yc}\footnote{One can find another form of the KSS bound with $D_{\text{shear}}=\eta/(s T)$, where $\eta$ is a shear viscosity, and $s$ the entropy density.}
\begin{align}\label{KSSB}
\begin{split}
D_{\text{shear}} \geq \frac{c^2}{4\pi} \, \tau_{\text{pl}} \,,
\end{split}
\end{align}
where $D_{\text{shear}}$ is the shear diffusion constant of the shear mode, $c$ the speed of light, and $\tau_{\text{pl}}:=\hbar/k_B T$ the Planckian time scale~\cite{Zaanen:2004aa,Sachdev:2011cs}.
The KSS bound \eqref{KSSB} first appeared to be confirmed by experimental data~\cite{Schafer:2009dj,Cremonini:2011iq,Luzum:2008cw,Nagle:2011uz,Shen:2011eg}, however it soon turns out that it can be easily violated by breaking a translational invariance~\cite{Alberte:2016xja,Hartnoll:2016tri,Burikham:2016roo,Rebhan:2011vd}.\footnote{The KSS bound has been further investigated in numerous ways including a broken rotational symmetry or the effects of anisotropy~\cite{Jain:2015txa,Rebhan:2011vd,Giataganas:2013lga,Jahnke:2014vwa,Inkof:2019gmh}, a higher derivative gravity \cite{Brigante:2007nu,Brigante:2008gz}. Moreover, recently, the relation between the momentum diffusivity and $\eta/s$ has been further elaborated in \cite{Baggioli:2020ljz}.}

\paragraph{Lower bound of $D$ with quantum chaos:} Inspired from the violation of the KSS bound, it has been motivated to find the bound of the diffusion constant ($D$) in terms of the velocity ($v$) and the time ($\tau$) scales:
\begin{align}\label{KSSB1}
\begin{split}
D \geq  v^2 \tau \,,
\end{split}
\end{align}
which is universal, for instance, not violated by breaking a translational invariance.

When the translational symmetry is broken, one may study the diffusive process governed by several diffusion constants: i) energy (or crystal) diffusion constant; \footnote{Depending on symmetry breaking patterns, one may study the energy diffusion constant (explicit symmetry breaking) or the crystal diffusion constant (spontaneous symmetry breaking). For the comprehensive review for this with the holographic toy model, see \cite{Baggioli:2021xuv} and references therein.} ii) charge diffusion constant.
Thus, it would be interesting to study which velocity and time scale can make a universal bound of such diffusion constants in the presence of broken translational symmetry.

It was proposed \cite{Blake:2016wvh,Blake:2016sud} that the relevant scales $(v, \tau)$ for a lower bound would be related to the properties from quantum chaos as
\begin{align}\label{KSSB2}
\begin{split}
v = v_{B} \,, \quad \tau = \tau_{L} \,,
\end{split}
\end{align}
where $v_B$ is the butterfly velocity and $\tau_L$ the Lyapunov time.
For the energy (or crystal) diffusive process, it was shown that the lower bound with quantum chaos \eqref{KSSB2} is robust and hard to break in many models~\cite{Lucas:2018wsc,Davison:2018ofp,Gu:2017njx,Ling:2017jik,Gu:2017ohj,Blake:2016jnn,Blake:2017qgd,Wu:2017mdl,Li:2019bgc,Ge:2017fix,Li:2017nxh,Ahn:2017kvc,Baggioli:2017ojd,Kim:2017dgz}.
Moreover, in recent years, the lower bound with \eqref{KSSB2} has also been understood with interesting phenomena from the ill-defined Green's function, called pole-skipping~\cite{Blake:2018leo,Jeong:2021zhz}.

On the other hand, in the case of the charge diffusion constant, it turned out that the bound with \eqref{KSSB2} does not hold and could easily be violated. For instance, one of the simplest ways to break the bound of the charge diffusion constant is by considering higher derivative couplings~\cite{Baggioli:2016pia}.\footnote{One can also see the violation of the lower bound for the charge diffusion in striped holographic matter~\cite{Lucas:2016yfl}, the SYK model~\cite{Davison:2016ngz}. In particular, the bound with \eqref{KSSB2} does not hold for the Gubser-Rocha model~\cite{Kim:2017dgz}, giving a divergence in the low $T$ limit ($m/T\gg1$). See also the case with massive gravity models~\cite{Amoretti:2014ola}.}

\paragraph{Upper bound of $D$ with the hydrodynamic convergence.}
In the similar perspective of the lower bound, in recent years, the holographic study of the upper bound has also been investigated as
\begin{align}\label{UPPB1}
\begin{split}
D \leq  v^2 \tau \,,
\end{split}
\end{align}
with the simple question: the diffusion constant will also be bounded from the above? 
Using the idea from the convergence of hydrodynamics, it is proposed \cite{Arean:2020eus} that the diffusion constant may have the upper bound with the following scale\footnote{More generally, it has been argued that the local equilibration time may set an upper bound on the diffusivity in \cite{Hartman:2017hhp,Lucas:2017ibu}.}:
\begin{align}\label{KSSB3}
\begin{split}
v = v_{\text{eq}} \,, \quad \tau = \tau_{\text{eq}} \,,
\end{split}
\end{align}
where $v_{\text{eq}}:=\omega_{\text{eq}}/k_{\text{eq}}$ is the equilibration velocity, $\tau_{\text{eq}}:=\omega_{\text{eq}}^{-1}$ the equilibration time.
In order to study the upper bound with ($v_{\text{eq}}, \tau_{\text{eq}}$) in \eqref{KSSB3}, we need to identify the equilibration scale ($\omega_{\text{eq}}, k_{\text{eq}}$). The equilibration scale is defined as the collision point in the ($\omega, k$) space between the diffusive hydrodynamic mode \eqref{eq2.11} and the first non-hydrodynamic mode. 

Note that the collision implies that, at such a scale, the dynamics of the system cannot be determined just by the hydrodynamic mode, i.e., the equilibration scale might be related to the radius of the convergence of hydrodynamic perturbative series~\cite{Withers:2018srf,Grozdanov:2019kge,Grozdanov:2019uhi,Heller:2020hnq,Heller:2020uuy,Heller:2013fn,Abbasi:2020ykq,Jansen:2020hfd,Grozdanov:2020koi,Choi:2020tdj,Grozdanov:2021gzh}.
Note also that \eqref{UPPB1} with \eqref{KSSB3} can be rewritten in terms of the equilibration scale ($\omega_{\text{eq}}, k_{\text{eq}}$) as
\begin{align}\label{UPPB2}
\begin{split}
D \leq  \frac{\omega_{\text{eq}}}{k_{\text{eq}}^{2}} \,,
\end{split}
\end{align}
where the upper bound (the equality) reflects the fact that the hydrodynamic dispersion at the quadratic order \eqref{eq2.11} becomes a good approximation at ($\omega_{\text{eq}}, k_{\text{eq}}$).

In this paper, our goal is to study the upper bound of the charge diffusion constant.
For the energy (or crystal) diffusive process, the upper bound proposal with \eqref{KSSB3} has been checked in \cite{Arean:2020eus,Wu:2021mkk,Jeong:2021zsv}.\footnote{In \cite{Arean:2020eus}, authors also checked that the upper bound proposal works for the shear diffusion constant in the translational invariant system.} However, the analysis for the upper bound proposal with the charge diffusion is still missing, thus we fill this gap in this paper. See  table \ref{SUM2} for the summary of the studies of the (lower/upper) bounds in holography. 
\begin{table}[]
\begin{tabular}{| C{4.42cm} | C{4.65cm} | C{4.68cm}  |}
\hline
   Diffusion   & Lower bound proposal  \eqref{KSSB2}  & Upper bound proposal \eqref{KSSB3}  \\ 
 \hline
 \hline
     \small{Energy (or Crystal) diffusion}   &  \small{Obeyed}      &  \small{Obeyed}    \\
 \hline
   \small{Charge diffusion}    &  \small{Violated}     &       ?          \\
 \hline
\end{tabular}
\caption{Summary of holographic studies for the bounds of diffusion constants. In this paper, we study the upper bound proposal \eqref{KSSB3} with the charge diffusion constant.}\label{SUM2}
\end{table}
Moreover, we also investigate the upper bound of the charge diffusion constant in the presence of the higher derivative coupling~\cite{Baggioli:2016pia} to examine the universality of the conjectured upper bound proposal~\eqref{KSSB3}.

This paper is organized as follows. 
In section \ref{sec2}, we introduce the gravity model to study the charge diffusion in holography. 
In section \ref{sec3}, we study the upper bound of the charge diffusion constant with the model presented in section \ref{sec2}.
In section \ref{sec4} we investigate the effect of the higher derivative coupling for the upper bound of the charge diffusion constant. 
Section \ref{sec6} is devoted to conclusions.

\section{Holographic setup}\label{sec2}

\subsection{Model}
We consider the Einstein-Maxwell-Axion model~\cite{Andrade:2013gsa} in (3+1) dimensions
\begin{align}\label{eq2.1}
S = \int d^4x \sqrt{-g} \left[R + \frac{6}{L^2} -\frac{1}{4}F^2 - \frac{1}{2}\sum^2_{i=1} (\partial \varphi_i)^2 \right]\,,
\end{align}
where we set the gravitational constant $16\pi G=1$ and we have included two matter fields: a U(1) gauge field $A$ with the field strength $F=\dd A$ and the massless scalar fields $\varphi_i = m x^i$ where $m$ denotes the strength of the translational symmetry breaking.

The model \eqref{eq2.1} allows the analytic background solution as
\begin{align}\label{METRICANST}
\begin{split}
\dd s^2 = & -f(r) \dd t^2 + \frac{\dd r^2}{f(r)} + \frac{r^2}{L^2} (\dd x^2 + \dd y^2)\,, \\
& f(r) = \frac{r^2}{L^2} \left(  1 + \frac{L^2 \mu^2 r_{h}^2}{4 r^4} - \frac{L^4 m^2}{2r^2} - \frac{r_{h}^3}{r^3} \left( 1 + \frac{L^2 \mu^2}{4 r_{h}^2} - \frac{L^4 m^2}{2 r_{h}^2} \right)  \right) \,, \\
A = & \, \mu \left(1-\frac{r_h}{r}\right) \dd t \,, \qquad \varphi_i = m x^i \,,
\end{split}
\end{align}
where $f(r)$ is the emblackening factor, $r_h$ is the black hole horizon, and $\mu$ denotes the chemical potential of the boundary field theory, $\mu = \lim\limits_{r\rightarrow\infty} A_t$. From here we set $L=1$.

The thermodynamic quantities from \eqref{METRICANST} read
\begin{align}\label{eq2.5}
T = \frac{f'(r_h)}{4 \pi} = \frac{1}{4 \pi}\left(3 r_h - \frac{\mu^2 + 2m^2}{4 r_h} \right) \,, \quad  \rho = \mu \, r_{h}\,,
\end{align}
where $\rho$ is interpreted as the expectation value of charge density.

\paragraph{Charge diffusion constant in holography:} Here we introduce the charge diffusion constant ($D_c$) as
\begin{align}\label{CDF1}
D_c := \frac{\sigma}{\chi} \,, \qquad \sigma = 1+\frac{\mu^2}{m^2} \,, \qquad \chi = \left(\frac{\partial\rho}{\partial\mu}\right)_{T} \,,
\end{align}
where $\sigma$ is the electric conductivity \cite{Andrade:2013gsa} and $\chi$ is the compressibility.
One can see that $m$ makes the conductivity finite as the reflection of the broken translational invariance.

At finite charge density, the generalized Einstein relation \cite{Hartnoll:2014lpa} shows the charge diffusion constant is coupled with the energy diffusion constant, i.e., in order to focus on the charge diffusion constant, we need to consider the zero charge density (or zero chemical potential) case. Therefore, we set $\mu=0$ in this paper.\footnote{At finite charge, the charge diffusion constant might be decoupled from the energy diffusion constant in the strong momentum relation limit~\cite{Kim:2017dgz}. In this paper, we only focus on the simplest case to study the charge diffusion constant: the $\mu=0$ case. We leave the study of the upper bound at finite charge in the strong momentum relaxation regime as future work.} The charge diffusion constant \eqref{CDF1} at zero charge can be expressed as 
\begin{align}\label{eq2.9}
    D_\text{c} = \frac{6}{4\pi T + \sqrt{6 m^2 + 16  \pi^2 T^2}}\,.
\end{align}
Note that the upper bound of the energy diffusion constant at $\mu=0$ has been investigated in \cite{Arean:2020eus}.

\subsection{Fluctuations for quasi-normal modes}
Based on the background solution \eqref{METRICANST} at zero charge, we consider the following gauge field fluctuations
\begin{align}\label{eq2.7}
\delta A_{t} (r, t, x, y) = \delta \bar{A}_{t}(r) \, e^{-i\omega t + i k x}\,,\quad 
\delta A_{x} (r, t, x, y) = \delta \bar{A}_{x}(r) \, e^{-i\omega t + i k x}\,,
\end{align}
which is relevant to the study of the hydrodynamic charge diffusion mode~\cite{Kovtun:2005ev}:
\begin{align}\label{eq2.11}
    \omega\, =\, -i\, D_\text{c}\, k^2\,,
\end{align}
where $D_c$ corresponds to \eqref{eq2.9}.

After Fourier transformation, one can find one single fluctuation equation of motion 
\begin{align}\label{eq2.13}
0=Z_A''(r) + \frac{r^3 \omega ^2 f'(r)-2 k^2 f(r)^2}{r f(r) \left(r^2
   \omega ^2-k^2 f(r)\right)}Z_A'(r) + \frac{r^2 \omega ^2-k^2
   f(r)}{r^2  f(r)^2}Z_A(r) \,,
\end{align}
by introducing the variable
\begin{align}\label{eq2.12}
Z_A(r)\, := k\, \delta \bar{A}_{t}(r) \, + \omega \, \delta\bar{A}_{x}(r) \,,
\end{align}
where it is composed of field fluctuations in \eqref{eq2.7}.

In order to study the quasi-normal mode spectrum for both the charge diffusion mode (the lowest mode) and the higher modes, we need to solve the equation of motion \eqref{eq2.13} with two boundary conditions: one from the horizon and the other from the AdS boundary.
Near the horizon, we impose the incoming boundary condition as 
\begin{align}\label{eq2.14}
Z_A = (r -r_h)^{\nu_{-}}\left(Z_A^{(I)} + Z_A^{(II)}(r-r_h)+ \cdots \right)\,,
\end{align}
where $\nu_{-} := - i \omega/4 \pi T$ and $Z_A^{(I, II)}$ are horizon coefficients.
Near the AdS boundary, the solution is expanded as
\begin{align}\label{GIV2}
Z_A = Z_A^{(S)} (1+\cdots) + Z_A^{(R)}r^{-1}(1+\cdots)\,,
\end{align}
where $Z_A^{(S)}$ is interpreted as the source term and $Z_A^{(R)}$ is for the response term via AdS/CFT dictionary.
Then, the quasi-normal mode spectrum can be found by the values of ($\omega, k$) where the source term $Z_A^{(S)}$ in \eqref{GIV2} is zero.

\section{The upper bound: Einstein-Maxwell-Axion model}\label{sec3}

In this section, we evaluate the quasi-normal modes from \eqref{eq2.13} and study the upper bound of the charge diffusion constant, \eqref{UPPB2}, with the equilibration scale ($\omega_{\text{eq}}, k_{\text{eq}}$).
In particular, we mainly focus on the low temperature ($T$) case because the upper bound is approached at low $T$. For the discussion beyond low $T$, see appendix \ref{sec5}.

\subsection{The equilibration scale}
The equilibration scale ($\omega_{\text{eq}}, k_{\text{eq}}$) is defined as the collision point in ($\omega, k$) space between the hydrodynamic mode \eqref{eq2.11} and the first non-hydrodynamic mode, i.e., the equilibration scale is related to the breakdown of the hydrodynamics.

\paragraph{Non-hydrodynamic modes (IR modes):} 
Let us first explain the non-hydrodynamic modes.
At low $T$, it is shown that the non-hydrodynamic modes correspond to the IR modes from IR green's function $\mathcal{G}_{\text{IR}}$~\cite{Arean:2020eus,Jeong:2021zsv}. 
We present $\mathcal{G}_{\text{IR}}$ of the fluctuation for the charge diffusion \eqref{eq2.12} as
\begin{align}\label{GIRCARGE}
    \mathcal{G_\text{IR}} = \left(\frac{3}{\pi}\right)^{1-2\Delta(0)} T^{1-2\Delta(0)} \frac{\Gamma\left(\frac{1}{2}-\Delta(0)\right)\Gamma\left(\Delta(0)-\frac{i \omega }{ 2\pi  T }\right)}{\Gamma\left(\Delta(0)-\frac{1}{2}\right) \Gamma\left(1-\Delta(0)-\frac{i \omega}{ 2 \pi  T}\right)} \,,
\end{align}
where $\Delta(0)$ is the scaling dimension of operator at the IR fixed point at zero wavevector.\footnote{We also present all the details for \eqref{GIRCARGE} in appendix \ref{APA} in a self-contained manner.}
Then, one can find the IR modes from the IR green's function \eqref{GIRCARGE} as
\begin{align}\label{IRMODES}
    \omega_n = -i\, 2 \pi T(n+\Delta(0)) \,,\quad n=0,1,2,\cdots \,,
\end{align}
where $\Delta(0)=1$ from ~\eqref{eq3.7}.

\paragraph{The equilibration scale ($\omega_{\text{eq}}, k_{\text{eq}}$):} 
In Fig. \ref{FIG2}, we display the representative result of quasi-normal modes at low $T$.
\begin{figure}[]
 \centering
     {\includegraphics[width=8cm]{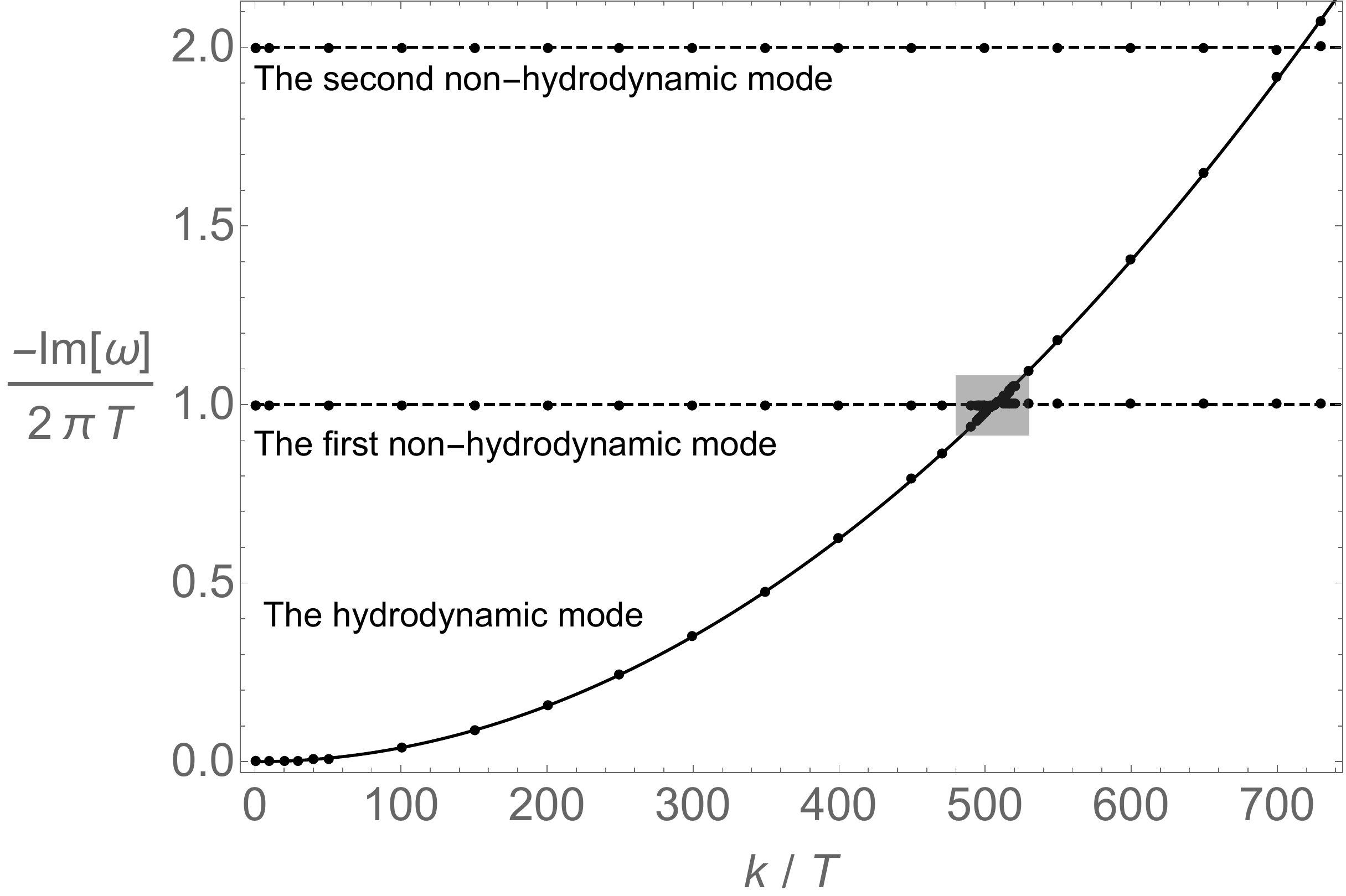} \label{FIG2}}
          \caption{Quasi-normal modes at $m/T = 10^5$. All dots represent quasi-normal modes. The solid line is the hydrodynamic mode \eqref{eq2.11}, and the dashed lines are IR modes \eqref{IRMODES}. In the gray square region, one can see the collision between the hydrodynamic mode and the first non-hydrodynamic mode.} \label{FIG2}
\end{figure}
From the figure, one can see that quasi-normal modes are well approximated with i) the hydrodynamic mode (solid line) \eqref{eq2.11}; ii) the IR modes (dashed line) \eqref{IRMODES}.
In particular, from the gray region in Fig. \ref{FIG2}, the equilibration scale ($\omega_{\text{eq}}, k_{\text{eq}}$) can be obtained from the collision $(\omega_{c}, k_{c})$ between the hydrodynamic mode and the first non-hydrodynamic mode.

The equilibration scale ($\omega_{\text{eq}}, k_{\text{eq}}$) is defined as the collision point $(\omega_{c}, k_{c})$ in absolute value
\begin{align}\label{absol}
    \omega_{\text{eq}}:= |\omega_\text{c}|\,,\quad k_{\text{eq}}:=|k_\text{c}|\,,
\end{align}
because the collision occurred in the complex ($\omega, k$) space for the cases with other diffusion constants~\cite{Arean:2020eus,Wu:2021mkk,Jeong:2021zsv}, satisfying 
\begin{align}\label{eq3.11}
    \omega_c \,=\, \omega_{\text{eq}} \, e^{i(\phi_k-\frac{\pi}{2})} \,, \quad k_\text{c} \,=\, k_{\text{eq}} \, e^{i\phi_k}\,,
\end{align}
where the finite phase $\phi_k$, $\phi_k \neq 0$, produces a complex $(\omega_{c}, k_{c})$.

\begin{figure}[]
 \centering
     \subfigure[Im $\omega$ vs k]
     {\includegraphics[width=7.2cm]{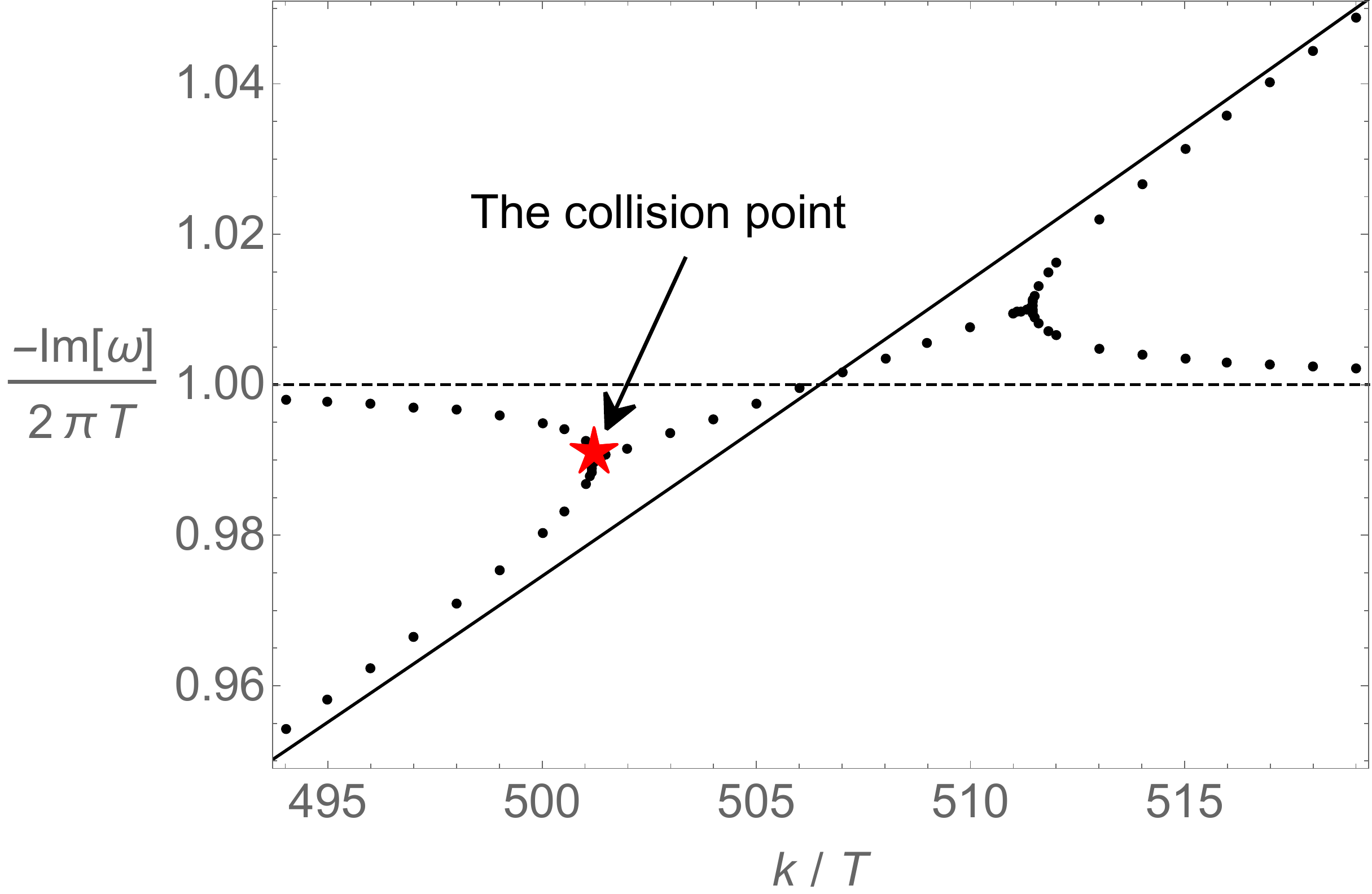} \label{FIG3a}}
     \subfigure[Re $\omega$ vs k]
     {\includegraphics[width=7.3cm]{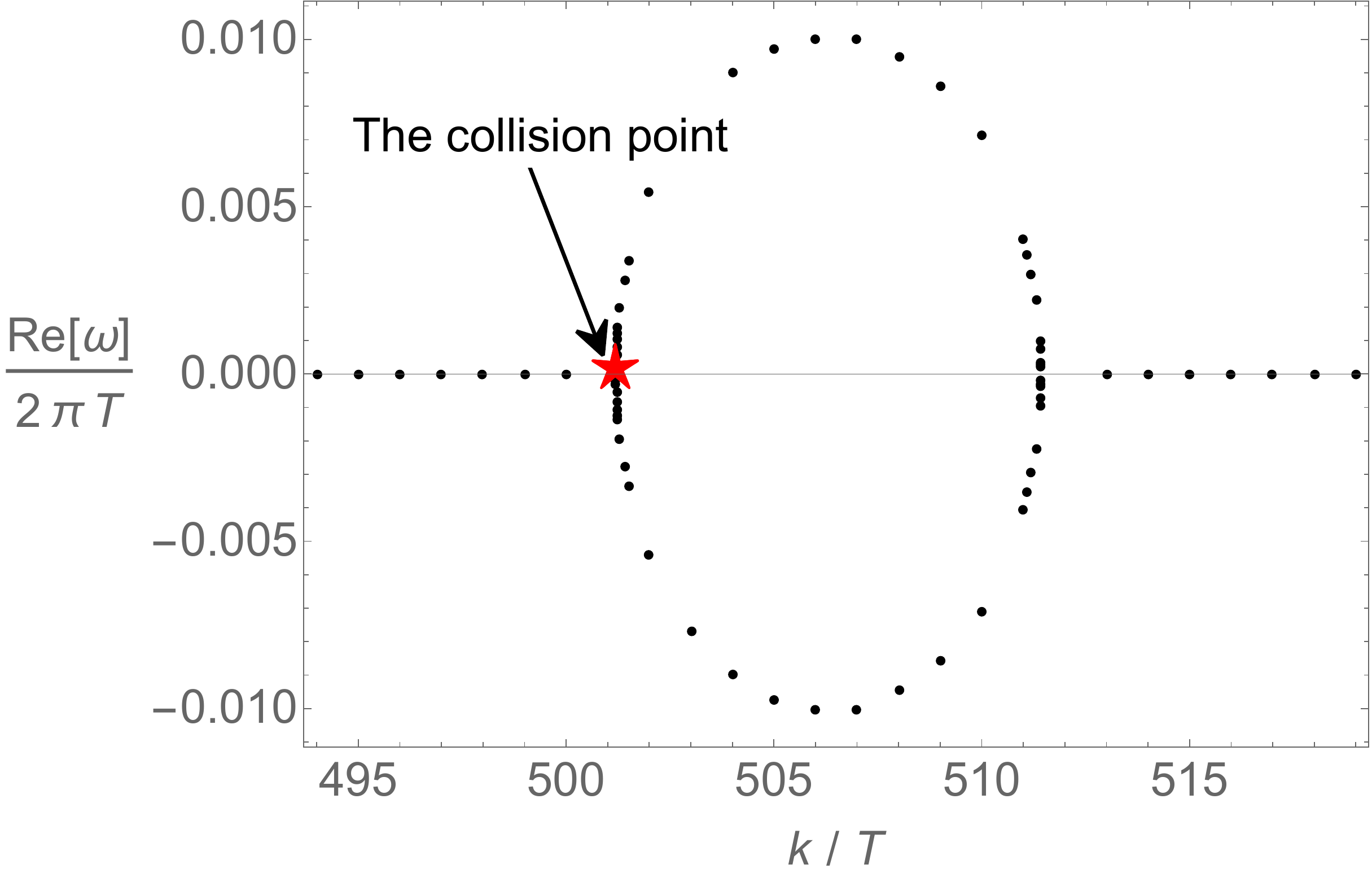} \label{FIG3b}}
          \caption{The collision between the hydrodynamic mode and the first non-hydrodynamic mode. The left figure is the zoom of the gray region in Fig. \ref{FIG2}. The red star corresponds to the collision point ($\omega_c, k_c$).} \label{FIG3}
\end{figure}

However, for the charge diffusion case, we found that the collision occurs in the real wavevector $k$, i.e., 
\begin{align}\label{}
 \phi_k = 0    \,,
\end{align}
which is a distinct feature not observed in other diffusion cases~\cite{Arean:2020eus,Wu:2021mkk,Jeong:2021zsv}. 
In Fig. \ref{FIG3}, we present how the hydrodynamic mode collides with the first non-hydrodynamic mode at real $k$. 
In Fig. \ref{FIG3a}, the collision point ($\omega_c, k_c$) is denoted as the red star. Note that, after the collision (red star), the quasi-normal modes would be a complex value with real $\omega$ and become pure imaginary at larger $k$. See Fig. \ref{FIG3b}.

\paragraph{The temperature dependence in ($\omega_{\text{eq}}, k_{\text{eq}}$):}
\begin{figure}[]
 \centering
     \subfigure[$\omega_{\text{eq}}$ vs $T$]
     {\includegraphics[width=7.1cm]{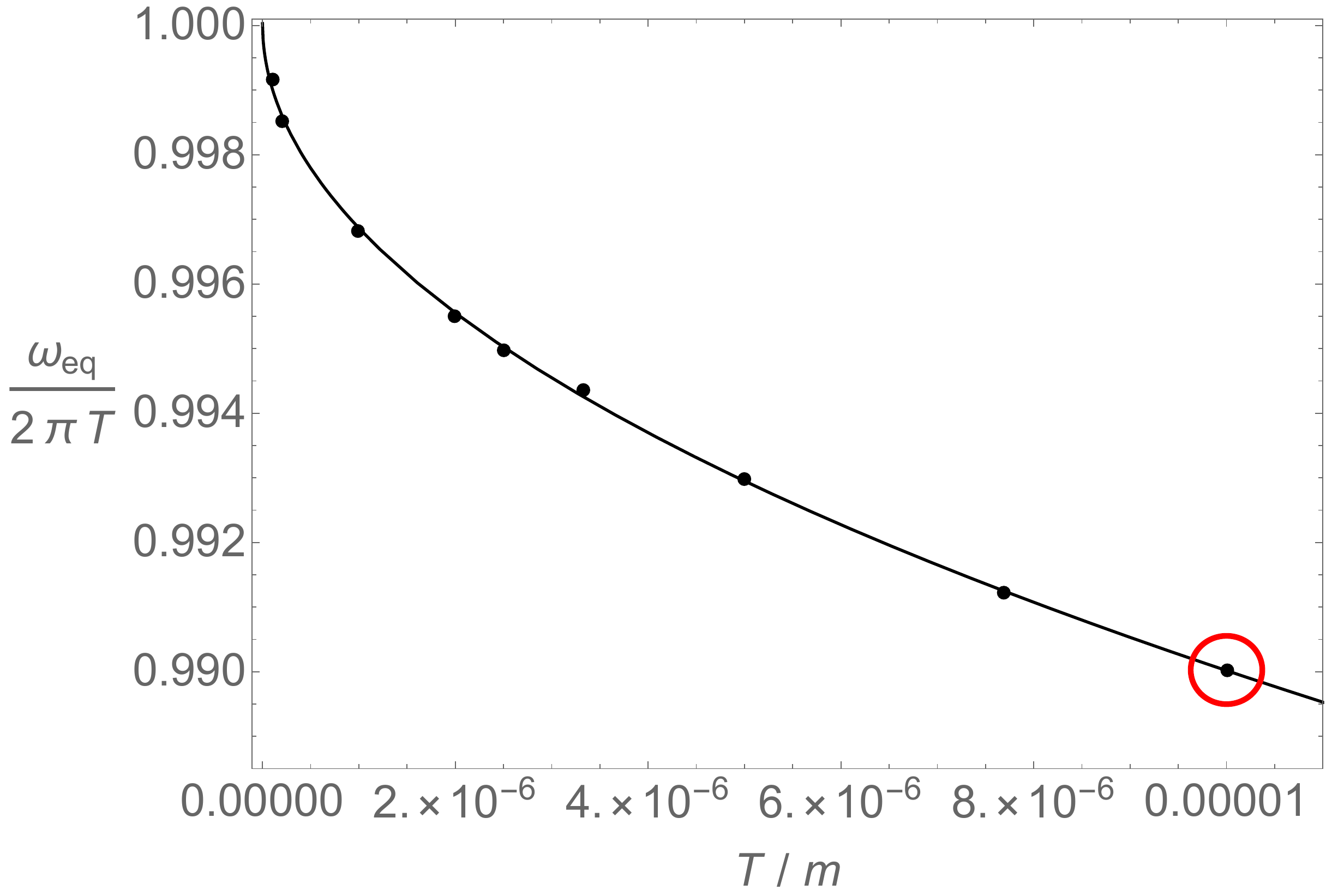} \label{FIG4a}}
     \subfigure[$k_{\text{eq}}$ vs $T$]
     {\includegraphics[width=6.8cm]{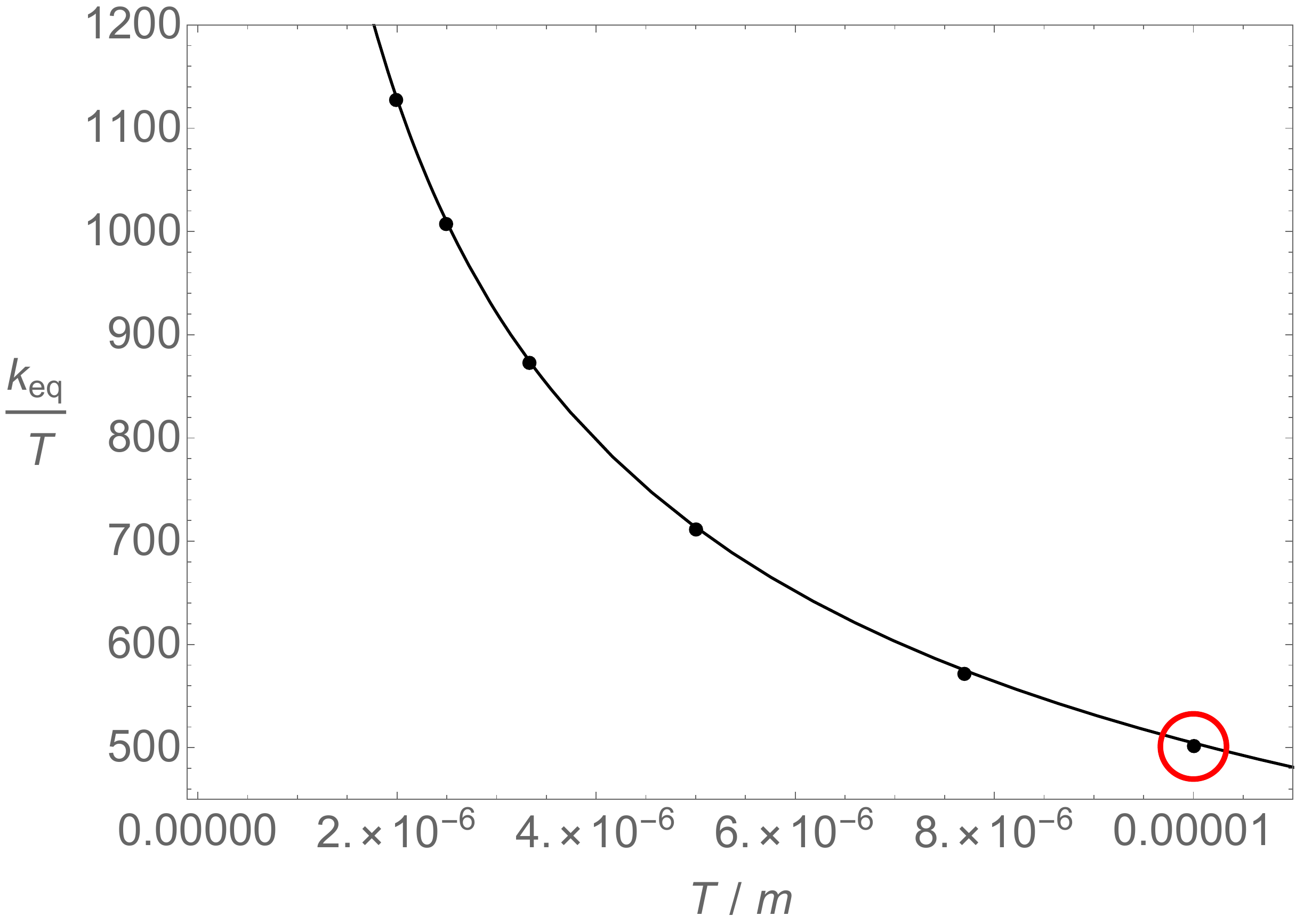} \label{FIG4b}}
          \caption{The temperature dependence of $\omega_{\text{eq}}$ and $k_{\text{eq}}$. Dots are numerical data and the solid lines are fitting curves \eqref{eq3.13}. The red circle corresponds to the collision point (red star) in Fig. \ref{FIG3}.} \label{FIG4}
\end{figure}
In Fig.~\ref{FIG4}, we show that the equilibration scale has the following temperature dependence as
\begin{align}\label{eq3.13}
    \frac{\omega_{\text{eq}}}{2 \pi T} \,\sim\, \Delta(0) \,+\, \# \sqrt{T} \,, \quad \frac{k_{\text{eq}}}{T} \,\sim\, \frac{\#}{\sqrt{T}}\,,
\end{align}
where $\Delta(0) = 1$ from \eqref{eq3.7}. Note that the red circle in Fig. \ref{FIG4} corresponds to the collision point (red star) in Fig. \ref{FIG3}.

\subsection{The upper bound}\label{sec3.3}

\paragraph{The upper bound of the charge diffusion constant ($D_c$):} 
With the conjectured upper bound proposal in \eqref{UPPB2}, now we study the upper bound of the charge diffusion constant ($D_c$):
\begin{align}\label{eq3.17}
    D_\text{c} \, \leq \, \frac{\omega_\text{eq}}{k^2_\text{eq}}\,,
\end{align}
where $D_c$ is \eqref{eq2.9} and the equilibration scale ($\omega_{\text{eq}}, k_{\text{eq}}$) is \eqref{eq3.13}.

\begin{figure}[]
 \centering
     {\includegraphics[width=7.2cm]{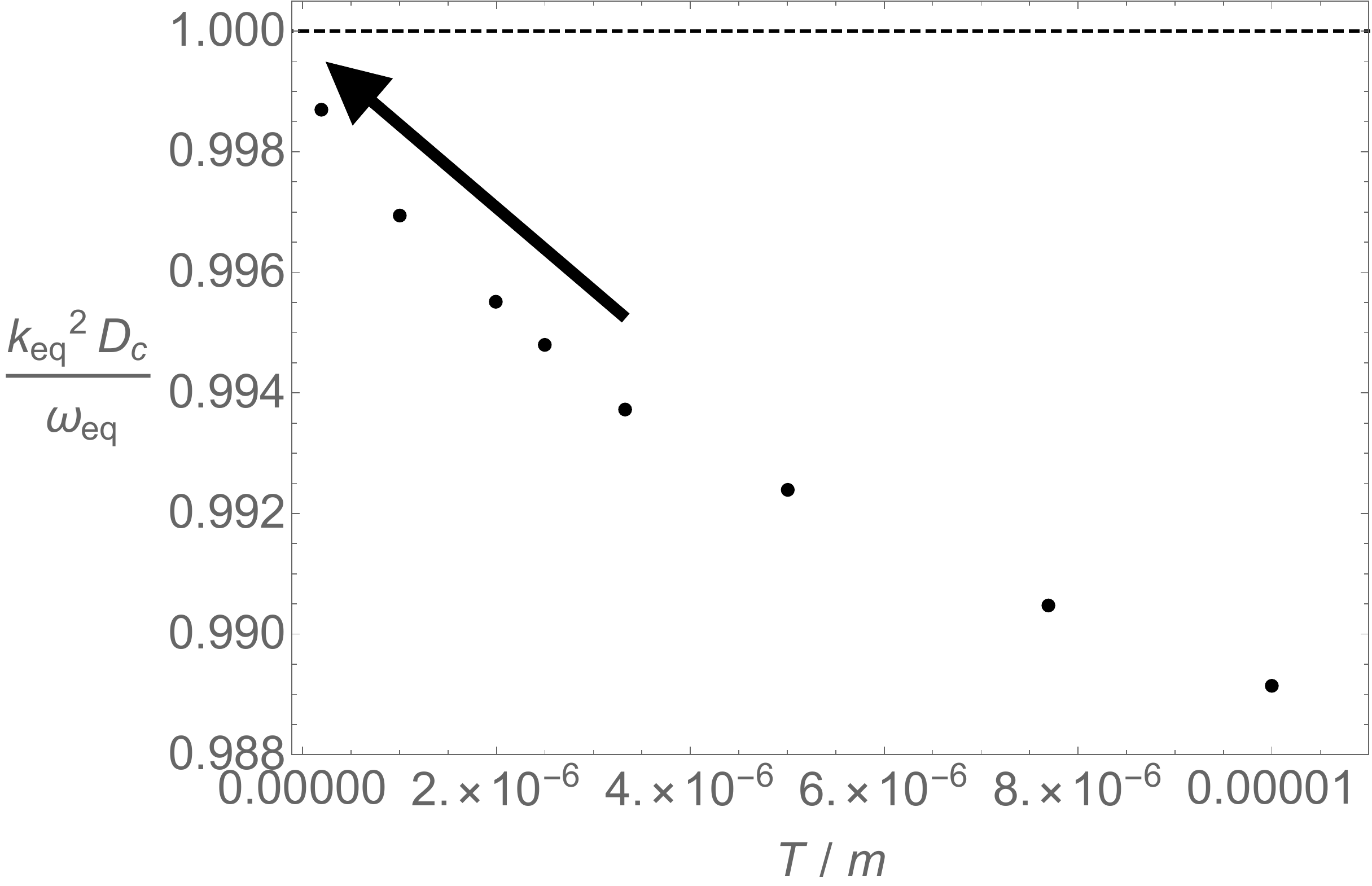}}
          \caption{The upper bound of the charge diffusion constant. The dashed line denotes the upper bound (or the equality) in \eqref{eq3.17}.} \label{FIG5}
\end{figure}

\begin{figure}[]
 \centering
     \subfigure[$m/T = 10^3$]
     {\includegraphics[width=7.2cm]{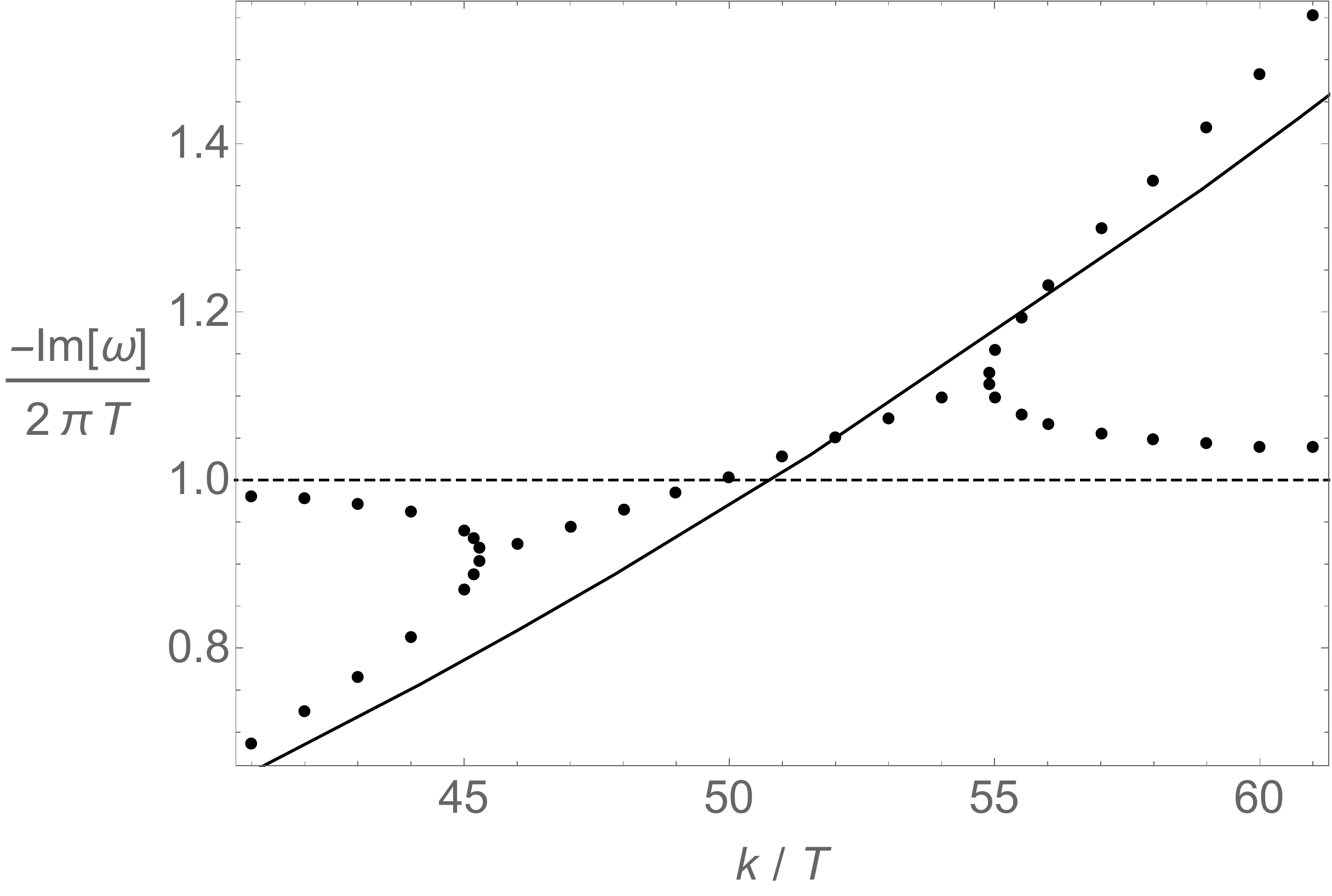} \label{FIG6a}}
     \subfigure[$m/T = 3 \times 10^5$]
     {\includegraphics[width=7.2cm]{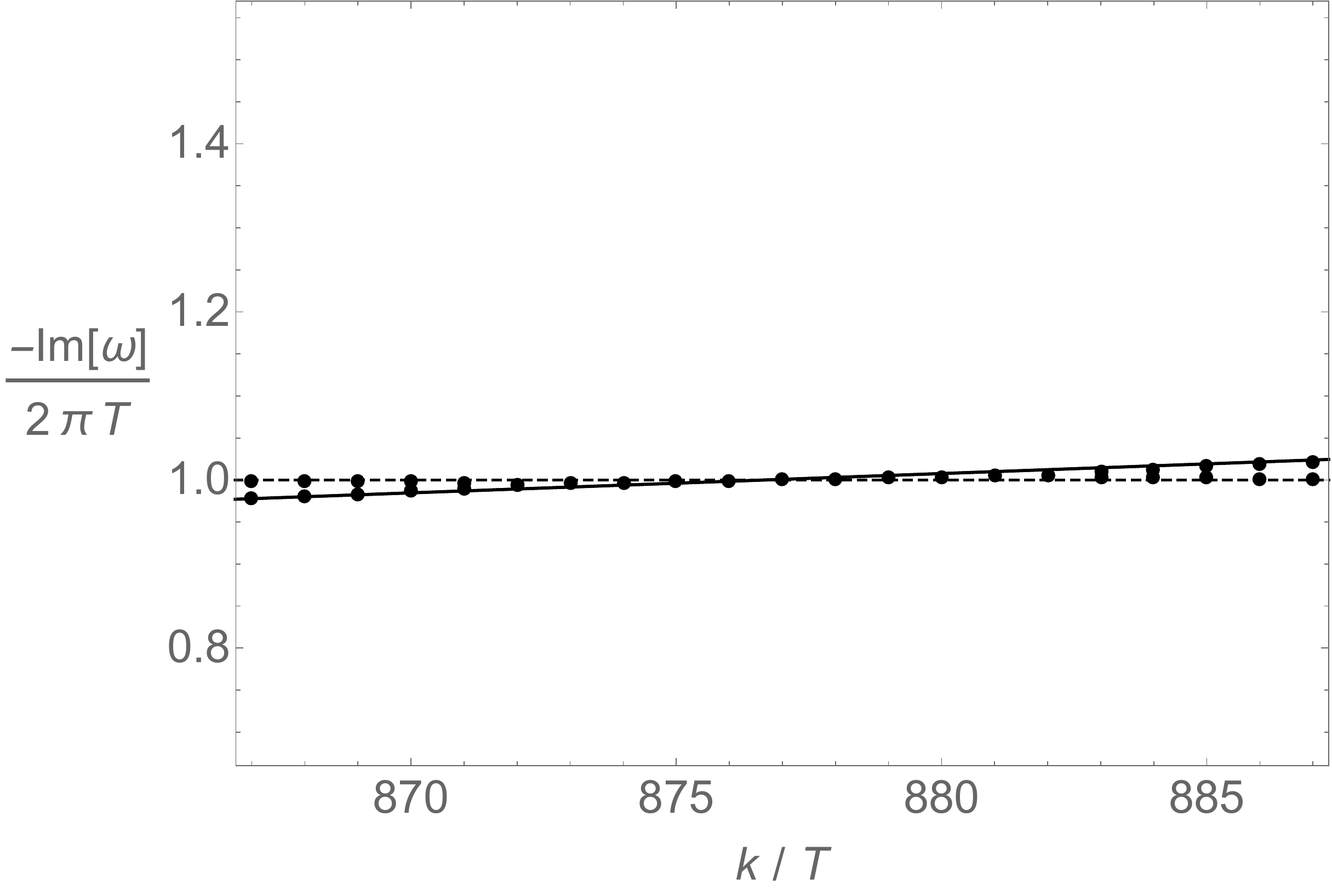} \label{FIG6b}}
          \caption{Quasi-normal modes with different $T$. As the temperature $T$ is lowered from Fig. \ref{FIG6a} to Fig. \ref{FIG6b}, quasi-normal modes (dots) are getting better and better approximated with hydrodynamic mode (solid line) \eqref{eq2.11} and IR modes (dashed line) \eqref{IRMODES}.} \label{FIG6}
\end{figure}

In Fig. \ref{FIG5}, we show that the conjectured upper bound proposal is valid for the case of the charge diffusion.
In other words, the equality in \eqref{eq3.17} is approached at low $T$. 
Note that the upper bound (or the equality) implies that, at low $T$, the quadratic hydrodynamic mode \eqref{eq2.11} becomes a good approximation even at ($\omega_{\text{eq}}, k_{\text{eq}}$), i.e.,
\begin{align}\label{eq3.16}
    1\,=\,\frac{k_\text{eq}^2\,D_\text{c}}{\omega_{\text{eq}}} \quad \rightarrow \quad \omega_\text{eq}\,=\,D_\text{c}\,k_\text{eq}^2 \,,
\end{align}
which also can be checked in Fig. \ref{FIG6}. 

\paragraph{Further comments on the upper bound of the diffusion constants ($D$):}

In the previous paragraph, we show that the upper bound proposal \eqref{UPPB2} is also valid for the charge diffusion constant in addition to other diffusion constants~\cite{Arean:2020eus,Wu:2021mkk,Jeong:2021zsv}. 
This may imply that there would be a universal feature for the upper bound proposal, which can be appeared with any diffusion constant $D$ at low $T$.
Thus, we may investigate further \eqref{UPPB2} with the low $T$ analysis to find such a universality.

First, let us consider the following low $T$ behavior of ($\omega_{\text{eq}}, k_{\text{eq}}$) with $D$ as
\begin{align}\label{eq3.19}
\begin{split}
    \omega_\text{eq} \,&=\, \left(\Delta(0)\,+\,c_1\,\left(\frac{T}{\Gamma}\right)^{p_1}\right)\,2\pi T \,, \qquad k_\text{eq}\,=\,\frac{c_2}{\left(\frac{T}{\Gamma}\right)^{p_2}}\,T\,,\cr
    D\,&=\,\frac{c_3}{\Gamma}\,+\,c_4\,\left(\frac{T^{p_3\,-\,1}}{\Gamma^{p_3}}\right)\,,
\end{split}
\end{align}
where $c_i$ are coefficients, $p_i$ the $T$-scaling power, and $\Gamma$ is an additional factor for the dimensionless analysis.\footnote{Note that dimensionless quantities would be ($\omega_{\text{eq}}/T, \, k_{\text{eq}}/T, \, D T$) together with $\Gamma/T$.}
For instance, in the case of $\Gamma=m$, the scaling power ($p_i$) would be
\begin{align}\label{numerator}
         \text{Energy diffusion:} & \quad p_1\,=\,1,\qquad \,\, p_2\,=\,\frac{1}{2},\qquad p_3\,=3\,, \cr
        \text{Charge diffusion:} & \quad p_1\,=\,\frac{1}{2},\qquad p_2\,=\,\frac{1}{2},\qquad p_3\,=2\,,    \end{align}
where the energy diffusion is from \cite{Arean:2020eus} and the charge diffusion case can be read from \eqref{eq3.13} with \eqref{eq2.9}.

With \eqref{eq3.19}, the upper bound of the diffusion constant, the equality in \eqref{UPPB2}, can be expanded at low $T$ as
\begin{align}\label{eq3.18}
\frac{k_\text{eq}^2 \,D}{\omega_\text{eq}} \,=\, \frac{(c_2)^2\,c_3}{2\pi \Delta(0)} \, \left(\frac{T}{\Gamma}\right)^{1\,-\,2p_2}  \,+\,  \dots \,,
\end{align}
where $\dots$ denotes sub-leading terms in $T$.
From this leading order result \eqref{eq3.18}, we may notice two things.
First, in order to have a universal upper bound independent of $T$, $p_2$ should be universal as 
\begin{align}\label{}
 \text{any diffusion constant $D$:} \quad p_2 = \frac{1}{2} \,,
\end{align}
for all $D$, which can also be observed in \eqref{numerator} as well as in other cases \cite{Arean:2020eus,Wu:2021mkk,Jeong:2021zsv}. 
Second, there would be a non-trivial relation between coefficients $c_i$ and $\Delta(0)$ to have the same upper bound, i.e.,
\begin{align}\label{LENTR}
 \frac{k_\text{eq}^2 \,D}{\omega_\text{eq}} \,=\, \frac{(c_2)^2\,c_3}{2\pi \Delta(0)} = 1 \,,
\end{align}
where the last equality is verified by the case of the charge diffusion as well as other diffusion cases~\cite{Arean:2020eus,Wu:2021mkk,Jeong:2021zsv}.\footnote{$c_3$ in \eqref{eq3.19} might be understood analytically, however $c_2$ mostly would be obtained from numerics by fitting so that the last equality in \eqref{LENTR} would be non-trivial.}
Note that, even for the case of the same diffusion constant $D$, $\Delta(0)$ would be different depending on $\Gamma$ so that one may not easily expect the relation in the last equality in \eqref{LENTR}:
for instance, for the energy diffusion, $\Delta(0)=2$ for $\Gamma=m$ ($m$ is an axion charge)~\cite{Arean:2020eus}, while $\Delta(0)=1$ for $\Gamma=H$ ($H$ is a magnetic field)~\cite{Jeong:2021zsv}.

\section{The upper bound with higher derivative coupling}\label{sec4}

In the previous section, we show that the conjectured upper bound also works for the charge diffusion constant, i.e., our work would be complementary to previous studies of upper bound of other diffusion constants~\cite{Arean:2020eus,Wu:2021mkk,Jeong:2021zsv}.

As demonstrated in the introduction, it was shown that the charge diffusion may not have a universal ``lower" bound in holography. One of the representative examples is a simple gravity model with a higher derivative coupling~\cite{Baggioli:2016pia}.

In this section, we further investigate the ``upper" bound of the charge diffusion constant in the presence of the higher derivative coupling to examine the universality of the conjectured upper bound proposal.

First, we review the higher derivative model and how the higher derivative coupling breaks the lower bound of the charge diffusion constant. Then we discuss the upper bound with the coupling.

\subsection{Model and the lower bound: a quick review}

\paragraph{Gauge-axion coupling model:} Let us consider the higher derivative coupling model~\cite{Baggioli:2016pia}
\begin{equation}
\begin{split}
S = \int d^4x \sqrt{-g} \left[ R + 6  - \frac{1}{4}F^2 - Tr[\mathcal{X}] - \frac{J}{4}Tr\left[\mathcal{X}F^2\right]      \,\right] \,, \label{ACTION}
\end{split}
\end{equation}
where
\begin{equation}\label{actionsetup}
{\mathcal{X}^\mu}_\nu\equiv \frac{1}{2}\sum_{i=1}^{2}\partial^\mu \varphi^i \partial_\nu \varphi^i  \,, \qquad Tr[\mathcal{X}\,F^2]\,\equiv\,[\mathcal{X}]^\mu_{\,\,\,\,\nu}~F^{\nu}_{\hspace{0.2cm}\nu'}~F^{\nu'}_{\hspace{0.2cm}\mu} \,.
\end{equation}
Note that \eqref{ACTION} becomes the Einstein-Maxwell-Axion model \eqref{eq2.1} at zero coupling, $J=0$.\footnote{The causality and the stability condition give the constraint to $J$ as $0\leqslant J\leqslant2/3$~\cite{Baggioli:2016pia}.}

The coupling $J$ does not affect the background equations of motion~\cite{Baggioli:2016pia} so that \eqref{ACTION} allows the same analytic solution \eqref{METRICANST} with the temperature \eqref{eq2.5}.

\paragraph{$D_{c}$ with the butterfly velocity $v_{B}$ (the lower bound):} The lower bound of the charge diffusion constant in \eqref{ACTION} can be studied with the following electric conductivity ($\sigma$), the butterfly velocity ($v_{B}$), and the Lyapunov time ($\tau_L$)
\begin{equation}\label{TDFD}
\begin{split}
\sigma \,=\, 1 - J \, \frac{m^2}{4 r_h^2} \,, \qquad v_{B}^2 \,=\, \frac{\pi T}{r_{h}}  \,, \qquad \tau_L \,=\, \frac{1}{2 \pi T} \,,
\end{split}
\end{equation}
at $\mu=0$.\footnote{For the details of \eqref{TDFD}, see \cite{Baggioli:2016pia}.} Recall that we need to consider the case at zero charge density for the study of the charge diffusion decoupled from the energy diffusion.

\begin{figure}[]
\centering
     {\includegraphics[width=7.0cm]{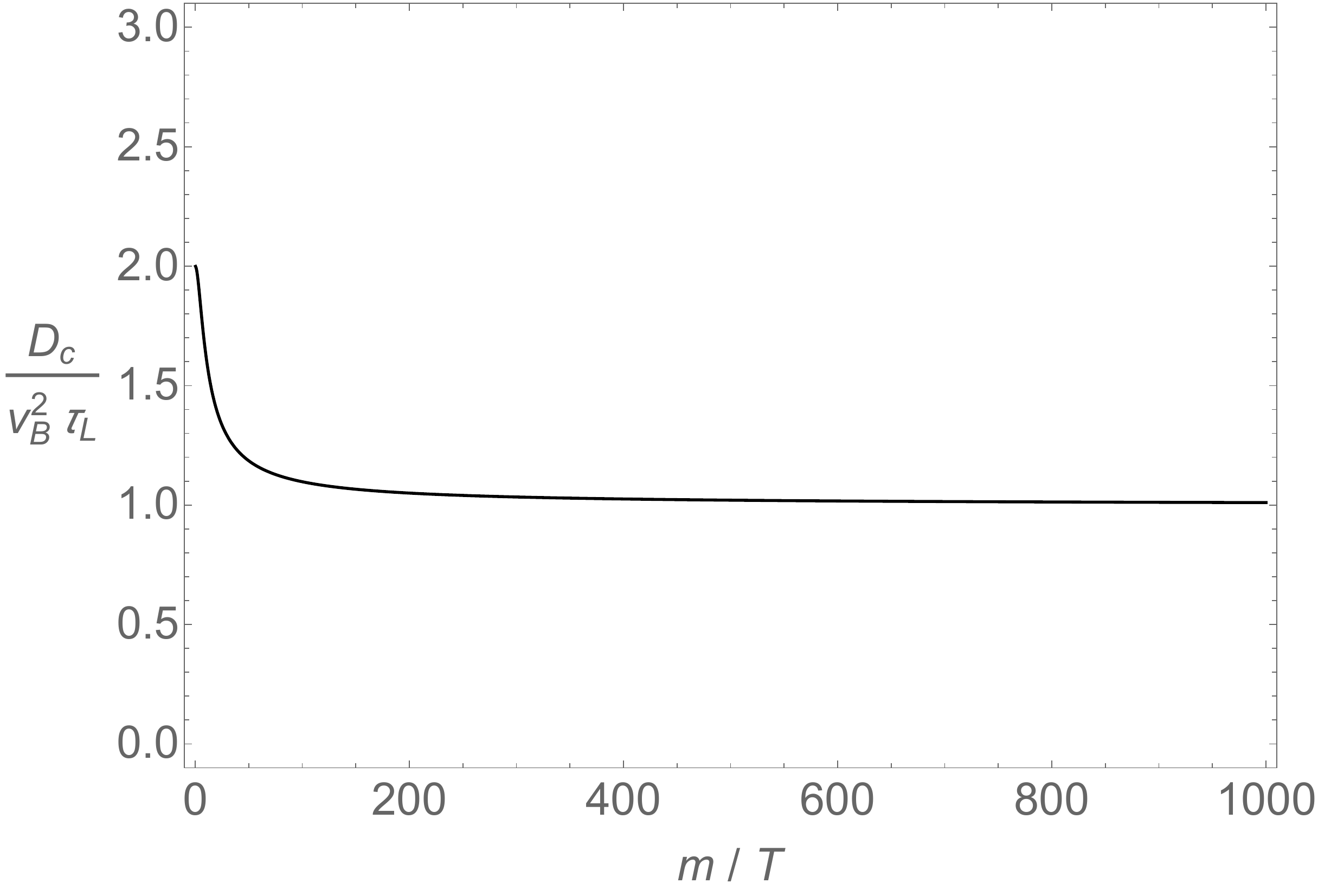}}
 \caption{The lower bound of $D_c$ at $J=1/3$, which approaches $\mathcal{B}_{L}=1$ in \eqref{LBCD} at low $T$ ($m/T\gg1$).}\label{STSKF2}
\end{figure}

Using \eqref{TDFD} with the definition of $D_c$ in \eqref{CDF1}, one can find that the charge diffusion constant and the butterfly velocity behave at low $T$ as
\begin{equation}\label{DDOEW}
\frac{D_{c}}{\tau_{L}} \,\sim\, \frac{\sqrt{6}\pi(2-3 J)}{m/T} \,,\qquad \frac{1}{v_{B}^2} \,\sim\,  \frac{1}{\sqrt{6}\pi}\frac{m}{T}\,,
\end{equation}
where the horizon $r_h$ is replaced by $T$ \eqref{eq2.5}.
Then, we have the lower bound of the charge diffusion constant, \eqref{KSSB2}, from \eqref{DDOEW} as
\begin{equation}\label{LBCD}
\mathcal{B}_{L} \,:=\, \frac{D_{c}}{v_{B}^2 \tau_{L}} \,=\, 2 - 3 J \,,
\end{equation}
where it is approached at low $T$. For instance, see the case of $J=1/3$ in Fig. \ref{STSKF2}.
From \eqref{LBCD}, we can see that the coupling $J$ can break the lower bound of the charge diffusion constant, i.e., $\mathcal{B}_{L}=0$ at $J=2/3$.

\subsection{The upper bound with the coupling}
Next, we study the upper bound of the charge diffusion, \eqref{KSSB3}, at a finite coupling $J$.
For this purpose, we need to investigate the coupling dependence in ($\tau_{\text{eq}}, v_{\text{eq}}$).

\paragraph{The equilibration time scale $\tau_{\text{eq}}$:}
Following the same procedure in appendix \ref{APA}, $\tau_{\text{eq}}$ can be read from the IR modes.
We found that the fluctuation equation of motion of \eqref{ACTION} is 
\begin{align}\label{IREOM22}
\begin{split}
0 \,=\, & \partial^2_\zeta Z_A + \left(\frac{2 \zeta }{\zeta ^2-\zeta_h^2}-\frac{(36-54J) \zeta 
   k^2}{9(2-3J) k^2 \left(\zeta^2-\zeta_h^2\right)-  \zeta_\omega ^2
   m^2}\right)\partial_\zeta Z_A  \\
   &+ \left(\frac{\zeta_\omega ^2}{9 \left(\zeta
   ^2-\zeta_h^2\right)^2}-\frac{(2-3J) k^2}{m^2 \left(\zeta ^2-\zeta
   _h^2\right)}\right)Z_A\,,
\end{split}
\end{align}
where the extremal geometry is \eqref{IRMETRIC}\footnote{Recall that the background geometry is not affected by the coupling $J$.} and also note that \eqref{IREOM22} becomes \eqref{IREOM} at $J=0$.

From \eqref{IREOM22}, one can see that the coupling is always coupled to the wavevector $k$. This means that $\tau_{\text{eq}}$ is independent of the coupling because $\omega_{\text{eq}}$ is evaluated at $k=0$ \eqref{eq3.10}, i.e.,
\begin{align}\label{veqvb1}
\begin{split}
\tau_{\text{eq}} := 1/\omega_{\text{eq}} = \frac{1}{2\pi T \Delta(0)} = \tau_L \,,
\end{split}
\end{align}
where \eqref{TDFD} is used in the last equality with $\Delta(0)=1$ \eqref{eq3.7}.

\begin{figure}[]
\centering
     {\includegraphics[width=7.0cm]{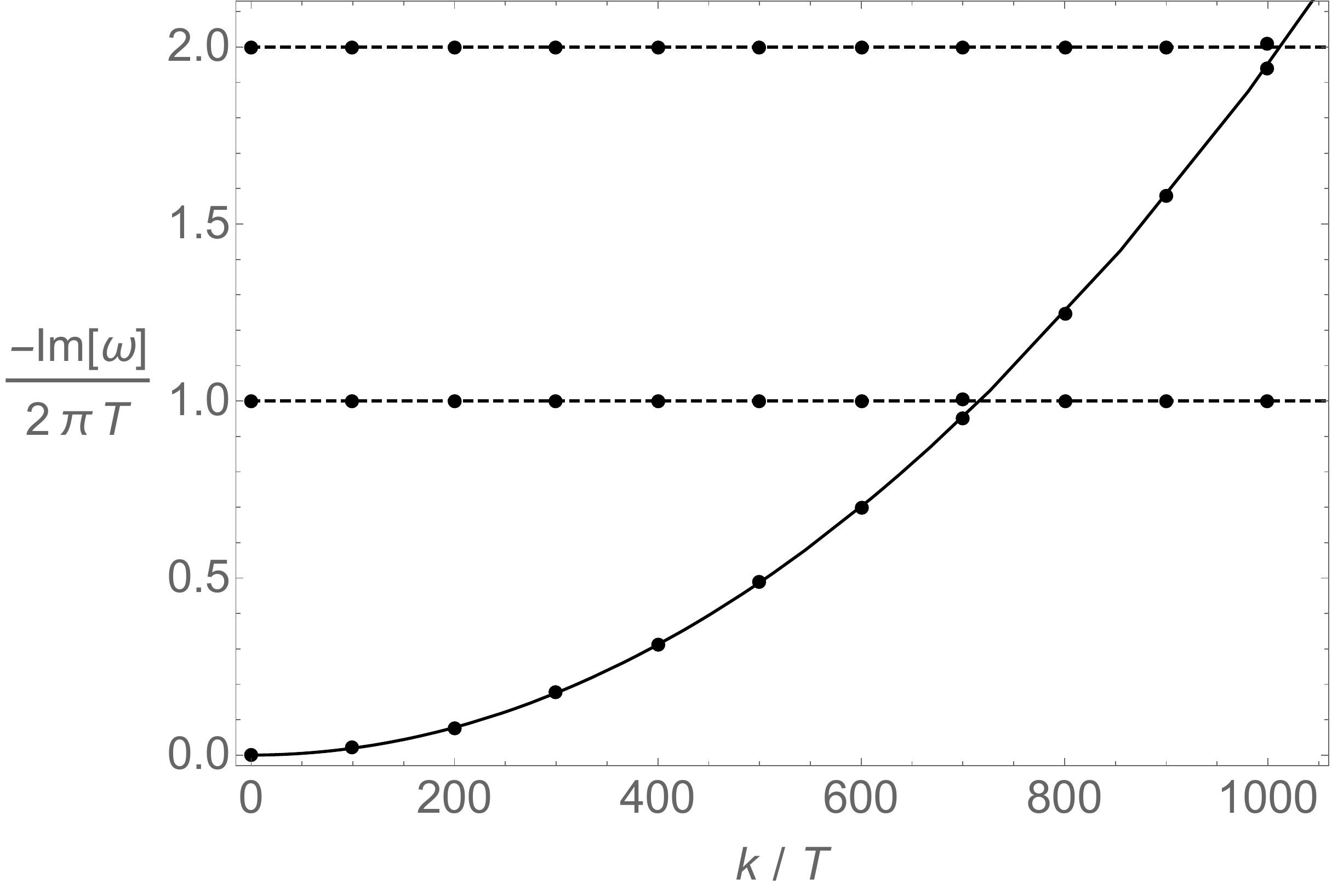}}
 \caption{Quasi-normal modes at $J=1/3$ with $m/T=10^5$. The solid line is the hydrodynamic mode~\eqref{eq2.11}. The dashed lines are IR modes~\eqref{IRMODES}.}\label{j1o3plots}
\end{figure}

\paragraph{The equilibration velocity scale  $v_{\text{eq}}$:} 
The velocity scale can be obtained as follows. 
\begin{align}\label{veqvb2}
\begin{split}
v_{\text{eq}}^2 \,:=\, \frac{\omega_{\text{eq}}^2}{k_{\text{eq}}^2} \,=\, \omega_{\text{eq}} \, D_c \,,
\end{split}
\end{align}
where we used \eqref{eq3.16} in the second equality. Note that it would be nontrivial if \eqref{eq3.16} is also valid at finite $J$: recall that \eqref{eq3.16} implies that, at low $T$, the quasi-normal modes are well approximated by both the hydrodynamic mode \eqref{eq2.11} and the IR modes \eqref{IRMODES}. We have numerically checked that \eqref{eq3.16} would also be valid at finite $J$.\footnote{As $J\rightarrow2/3$, we checked that $D_c$ is decreasing \eqref{DDOEW}, while $k_{\text{eq}}$ increases so that \eqref{eq3.16} would be respected at $J\rightarrow2/3$.} For instance, see the representative example for $J=1/3$ in Fig. \ref{j1o3plots}.\footnote{\eqref{eq3.16} at finite $J$ is already implying that the upper bound proposal is valid at finite coupling.} 
With the time scale \eqref{veqvb1}, the equilibration velocity \eqref{veqvb2} can be further expressed as
\begin{align}\label{vewdcrl}
\begin{split}
v_{\text{eq}}^2 \,=\, \omega_{\text{eq}} \, D_c \,=\, \frac{D_c}{\tau_L} \,,
\end{split}
\end{align}
and, using \eqref{DDOEW}, one can find the relation between the equilibration velocity and the butterfly velocity as
%
\begin{align}\label{veqvb3}
\begin{split}
v_{\text{eq}}^2  \,=\, (2-3J) \,v_{B}^2 \,.
\end{split}
\end{align}
Thus, the coupling $J$ does not affect $\tau_{\text{eq}}$ \eqref{veqvb1}, but $v_{\text{eq}}$ \eqref{veqvb3}.

\paragraph{$D_{c}$ with the equilibration velocity $v_{\text{eq}}$ (the upper bound):} 
Then, now one can find the upper bound of the charge diffusion at finite $J$ as
\begin{equation}\label{LBCD2222}
\mathcal{B}_{U} \,:=\, \frac{D_{c}}{v_{\text{eq}}^2 \tau_{\text{eq}}} \,=\,  \frac{D_{c}}{v_{B}^2 \tau_{L}} \, \frac{1}{2-3J} \,=:\, \mathcal{B}_{L}  \, \frac{1}{2-3J} \,,
\end{equation}
where we used \eqref{veqvb1} and \eqref{veqvb3}.
One may think that the upper bound ($\mathcal{B}_{U}$) in \eqref{LBCD2222} is diverging at $J=2/3$ so that the coupling also breaks the upper bound as well as the lower bound ($\mathcal{B}_{L}$) in \eqref{LBCD}. However, the upper bound would be still valid at finite coupling $J$, because $\mathcal{B}_{L}$ in \eqref{LBCD} cancels out the coupling dependence, $1/(2-3J)$, in \eqref{LBCD2222}, in other words, we have $\mathcal{B}_{U} = 1$ for all $J$.
%


\section{Conclusion}\label{sec6}

We have studied the upper bound of the charge diffusion constant with the conjectured upper bound proposal~\cite{Arean:2020eus,Wu:2021mkk,Jeong:2021zsv}:
\begin{align}\label{CONF1}
    D \, \leq \, \frac{\omega_\text{eq}}{k^2_\text{eq}}\,,
\end{align}
where the equilibration scale ($\omega_{\text{eq}}, k_{\text{eq}}$) is identified with the collision point between the diffusive hydrodynamic mode and the first non-hydrodynamic mode. 
The upper bound (an equality) in \eqref{CONF1} is approached at low $T$ with
\begin{align}\label{CONF2}
    \quad  \omega_\text{eq} \rightarrow 2\pi T \Delta(0) \,, \quad k_{\text{eq}}^2 \rightarrow \frac{\omega_{\text{eq}}}{D}\,,
\end{align}
which implies that the hydrodynamic mode at quadratic order \eqref{eq2.11} would be a good approximation even around the equilibration scale.

\paragraph{Charge diffusion constant and the upper bound:} 
We check that the upper bound proposal \eqref{CONF1} also works for the charge diffusion constant, $D_c$, in addition to other diffusion constants: energy diffusion constant~\cite{Arean:2020eus,Jeong:2021zsv}, shear diffusion constant~\cite{Arean:2020eus}, crystal diffusion constant~\cite{Wu:2021mkk,Jeong:2021zsv}.
This implies that there would be a universal property for the upper bound, independent of the type of diffusion constants. From the low temperature ($T$) analysis in \eqref{eq3.19}, we found that $k_{\text{eq}}$ plays the major role in the upper bound of any diffusion constant $D$, as
\begin{align}\label{}
    \frac{k_\text{eq}^2 \,D}{\omega_\text{eq}} \sim T^{1-2 p_2}  \,,
\end{align}
where $p_2$ is from $k_{\text{eq}}\sim T^{1-p_2}$. Thus, one can notice that $p_2$ should be universal as $p_2 = 1/2$ in order to have the upper bound (or a $T$-independent quantity).

In addition to the universal feature, $p_2 =1/2$, we also found that the charge diffusion case would have a distinct feature in the phase of $k$, $\phi_k$, in \eqref{eq3.11}, which is not observed in other diffusion constants:
\begin{align}\label{CONF3}
\begin{split}
\text{Charge diffusion}: \quad \phi_k = 0 \,, \qquad 
\text{Other diffusions}~: \quad \phi_k \neq 0 \,.
\end{split}    
\end{align}
For instance, energy diffusion constant has a finite $T$-dependence as $\phi_k\sim T^{7/2-\Delta(0)}$~\cite{Jeong:2021zsv}.
Note that \eqref{CONF3} implies the collision occurs at a real wavevector for the charge diffusion while it is complex for other diffusions~\cite{Arean:2020eus,Wu:2021mkk,Jeong:2021zsv}.\footnote{In \cite{Baggioli:2020loj}, it was shown that the radius of convergence of linear hydrodynamics in liquids is also related to the real wavevector, called k-gap \cite{Baggioli:2019jcm}.}

It is investigated that the quantum chaos property would be related with the pole-skipping phenomena in the ``upper"-half-frequency-plane~\cite{Blake:2018leo}. However, the pole-skipping point of the charge diffusion is in the ``lower"-half-frequency-plane~\cite{Blake:2019otz,wipCW} so that the charge diffusion may not be related to quantum chaos.
The real wavevector collision in \eqref{CONF3}, $\phi_k = 0$, might be a new piece of supporting evidence for this. Note that one of the signals of the quantum chaos is the level repulsion from random matrix theory~\cite{Bohigas:1984aa,BTZ-Poisson,BERRY1981163}, i.e., the quantum chaos property might be related to the case with $\phi_k \neq 0$ in our language.

\paragraph{Universal upper bound:} 
We further investigated the upper bound of the charge diffusion constant with the higher derivative couplings and found that, unlike the lower bound \eqref{LBCD}, the coupling cannot break the upper bound of the charge diffusion constant \eqref{LBCD2222}, thus we speculate that the conjectured upper bound
\begin{align}\label{CONF4}
    \frac{k_\text{eq}^2 \,D_c}{\omega_\text{eq}} = 1  \,,
\end{align}
would be universal independent of the couplings (or UV data), but only depends on the IR fixed points as in the universal lower bound of the energy diffusion constant.\footnote{We are grateful to Yan Liu and Xin-Meng Wu for sharing the preliminary results for the case of the charge diffusion in the conformal to AdS$_2$ fixed point~\cite{wipYL}: it seems that the gauge coupling may play an important role to study the upper bound as well the IR geometry. For the detailed discussion and extension, we refer to their forthcoming work~\cite{wipYL}. }
It would be interesting to investigate how \eqref{CONF4} (as well as the case with other diffusion constants) can be generalized to generic IR fixed points.\footnote{Recall that \eqref{CONF4} is for the AdS$_2$ fixed point. We expect that there could be a universal constant on the right hand side of \eqref{CONF4} because the leading constant at low $T$ does not depend on the UV data $(T, \Gamma)$ as can be seen in \eqref{LENTR}.}
We leave this subject as future work and hope to address it in the near future.

\acknowledgments

We would like to thank  {Yongjun Ahn, Matteo Baggioli, Wei-Jia Li, Yan Liu, Xin-Meng Wu}  for valuable discussions and correspondence. We thank {Yan Liu, Xin-Meng Wu} for sharing with us unpublished results.
This work was supported by the National Key R$\&$D Program of China (Grant No. 2018FYA0305800), Project 12035016 supported by National Natural Science Foundation of China, the Strategic Priority Research Program of Chinese Academy of Sciences, Grant No. XDB28000000, Basic Science Research Program through the National Research Foundation of Korea (NRF) funded by the Ministry of Science, ICT $\&$ Future Planning (NRF- 2021R1A2C1006791) and GIST Research Institute (GRI) grant funded by the GIST in 2021.
K.-B. Huh and H.-S. Jeong contributed equally to this paper and should be considered co-first authors.

\appendix


\section{Non-hydrodynamic poles: Infra-red modes}\label{APA}

It was shown that non-hydrodynamic modes are associated with infra-red Green's function~\cite{Arean:2020eus,Jeong:2021zsv}.
The infra-red Green's function can be calculated by solving the fluctuation equation in the extremal geometry with scaling dimension $\Delta(k)$ at the infra-red fixed point. For the details, we refer the readers to \cite{Jeong:2021zsv}.

\paragraph{The extremal geometry:} 
From ~\eqref{eq2.5}, we can consider the following relation at zero temperature: 
\begin{align}\label{zeroT}
r_h = r_e \,, \quad r_e = \frac{m}{\sqrt{6}}  \,.
\end{align}
Moreover, considering the following coordinate transformation, one can have the extremal geometry
\begin{align}\label{eq3.2}
r\, =\, r_e\, +\, \epsilon\, \zeta\,, \quad r_h\, =\, r_e\, + \epsilon\, \zeta_h\,, \quad t\, =\, \frac{u}{\epsilon}  \,,
\end{align}
where $\zeta_h \,=\, 4\, \pi\, \delta T\, /\, 9$ represents a small temperature correction.

Considering ~\eqref{eq3.2} in the $\epsilon \,\rightarrow\,0$ limit, the metric of ~\eqref{METRICANST} is transformed into the extremal geometry as
\begin{align}\label{IRMETRIC}
\dd s^2 = -\frac{\zeta^2}{L_{2}^2}\left(1-\frac{\zeta_h}{\zeta}\right)^2 \dd u^2 + \frac{L_{2}^2}{\zeta^2\left(1-\frac{\zeta_h}{\zeta}\right)^2}\dd r^2 + r_e^2\left(\dd x^2+\dd y^2\right) \,,
\end{align}
where the AdS$_{2}$ radius $L_{2}$ is $\sqrt{1/6}$.

\paragraph{Fluctuation equation for IR modes:} 
Similarly, we can do the coordinate transformation in the Fourier space as
\begin{align}\label{TR1}
r \,=\, r_e\, +\, \epsilon\, \zeta\,, \quad r_h\, =\, r_e\, +\, \epsilon\, \zeta_h\,, \quad \omega\, =\, \epsilon\, \zeta_\omega  \,,
\end{align}
and from this transformation we can express the fluctuation equation ~\eqref{eq2.13} in the extremal geometry as
\begin{align}\label{IREOM}
\begin{split}
0 \,=\, &\partial^2_\zeta Z_A+\left(\frac{2 \zeta }{\zeta ^2-\zeta_h^2}-\frac{36 \zeta 
   k^2}{18 k^2 \left(\zeta^2-\zeta_h^2\right)-  \zeta_\omega ^2
   m^2}\right)\partial_\zeta Z_A \\
   &+\left(\frac{\zeta_\omega ^2}{9 \left(\zeta
   ^2-\zeta_h^2\right)^2}-\frac{2 k^2}{m^2 \left(\zeta ^2-\zeta
   _h^2\right)}\right)Z_A\,,
\end{split}
\end{align}
where the second term of the first derivative of \eqref{IREOM}, the $k$-dependent term, did not appear in other diffusion cases~\cite{Arean:2020eus,Wu:2021mkk,Jeong:2021zsv}.

In the $\text{AdS}_\text{2}$ boundary $(\zeta \rightarrow \infty)$, the solution for \eqref{IREOM} can be expanded:
\begin{align}\label{eq3.6}
    Z_A = Z^{(S)}\zeta^{\Delta(k) - 1} + Z^{(R)}\zeta^{-\Delta(k)} \,,
\end{align}
where $Z_A^{(S)}$ is interpreted as the source term and $Z_A^{(R)}$ is for the response term and $\Delta(k)$ is the operator of dimension at the infra-red fixed point
\begin{align}\label{eq3.7}
    \Delta(k) = \frac{1}{2} + \frac{\sqrt{8 k^2 + m^2}}{2m} \,.
\end{align}

\paragraph{The non-hydrodynamic modes:} 
According to the holographic dictionary, the infra-red Green's function, $\mathcal{G_\text{IR}}$, can be calculated by solving the fluctuation equation ~\eqref{IREOM} near the $\text{AdS}_\text{2}$ boundary $(\zeta \rightarrow \infty)$~\cite{Faulkner:2009wj,Hartnoll:2012rj}:
\begin{align}\label{eq3.8}
    \mathcal{G_\text{IR}} \varpropto \frac{Z^{(R)}}{Z^{(S)}} \,,
\end{align}
where $Z^{(R)}$ and $Z^{(S)}$ are the coefficients of the solution ~\eqref{eq3.6}, respectively. 

\begin{figure}[]
 \centering
     {\includegraphics[width=7.2cm]{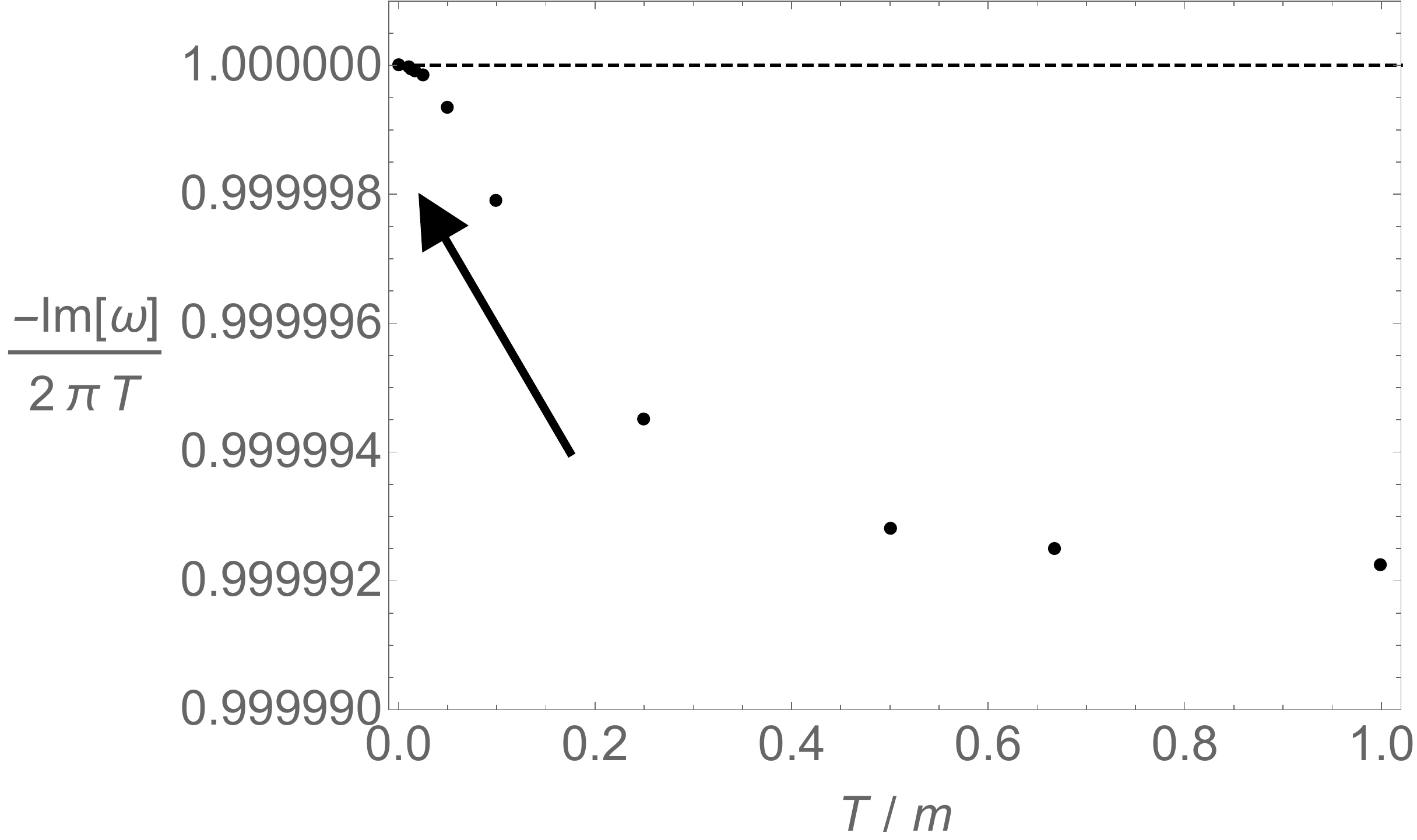}} 
          \caption{The first-non hydrodynamic mode at $k/T = 1/100$. Poles (black dots) are approaching the IR mode (dashed line) \eqref{eq3.10}, the first IR mode  ($\omega_0$), at low $T$.} \label{FIG0}
\end{figure}

Then one can find $\mathcal{G}_{\text{IR}}$ at $k=0$\footnote{We focus on the case at $k = 0$, which is sufficient for the non-hydrodynamic mode~\cite{Arean:2020eus,Jeong:2021zsv}.} as
\begin{align}\label{eq3.9}
    \mathcal{G_\text{IR}} = \left(\frac{3}{\pi}\right)^{1-2\Delta(0)} T^{1-2\Delta(0)} \frac{\Gamma\left(\frac{1}{2}-\Delta(0)\right)\Gamma\left(\Delta(0)-\frac{i \omega }{ 2\pi  T }\right)}{\Gamma\left(\Delta(0)-\frac{1}{2}\right) \Gamma\left(1-\Delta(0)-\frac{i \omega}{ 2 \pi  T}\right)} \,,
\end{align}
where the pole of the $\mathcal{G}_{\text{IR}}$ of \eqref{eq3.9}, IR modes, are
\begin{align}\label{eq3.10}
    \omega_n = -i\, 2 \pi T(n+\Delta(0)) \,,\quad n=0,1,2,\cdots \,,
\end{align}
with $\Delta(0)=1$ from ~\eqref{eq3.7}. 
In Fig.~\ref{FIG0}, we show that the non-hydrodynamic pole is approaching the IR modes \eqref{eq3.10} at low $T$.

\section{Quasi-normal modes beyond low temperature}\label{sec5}

In Sec.~\ref{sec3}, we mainly focused on the ``low" temperature regime ($m/T\gg1$) in order to study the bound of the charge diffusion constant, showing that, at low $T$, the quasi-normal modes are well approximated with i) the hydrodynamic mode \eqref{eq2.11}; ii) the IR modes \eqref{eq3.10}. See Fig. \ref{FIG2}.

\paragraph{Motivation:} 
In this section, we further investigate the quasi-normal modes at ``higher" temperature (i.e., smaller $m/T$) to show how \eqref{eq2.11} and \eqref{eq3.10} would be good approximations to the quasi-normal modes at low $T$.
To our knowledge, our work is the first holographic study showing the excellent applicability of approximations (\eqref{eq2.11}, \eqref{eq3.10}) in terms of the full dynamics of quasi-normal modes from high $T$ to low $T$.\footnote{In order to deliver the main message clearly, we only display the imaginary part of the quasi-normal modes in the paper.}
Although we focused on the charge diffusion constant in this paper, we believe similar dynamics may occur in the case of other diffusion constants~\cite{Arean:2020eus,Wu:2021mkk,Jeong:2021zsv}.\footnote{For the energy diffusion case, the collision between the hydrodynamic mode and the non-hydrodynamic mode appears at complex $k$ at low $T$~\cite{Arean:2020eus,Jeong:2021zsv} and real $k$ at high $T$~\cite{Davison:2014lua,Jeong:2021zhz}. On the other hand, for the charge diffusion, the collision occurs at real $k$ for all $T$ as we will show below. Therefore, dynamics from the energy diffusion would be more complicated than the charge diffusion case.}

\begin{figure}[]
 \centering
     \subfigure[$m/T = 10$ (high $T$)]
     {\includegraphics[width=4.8cm]{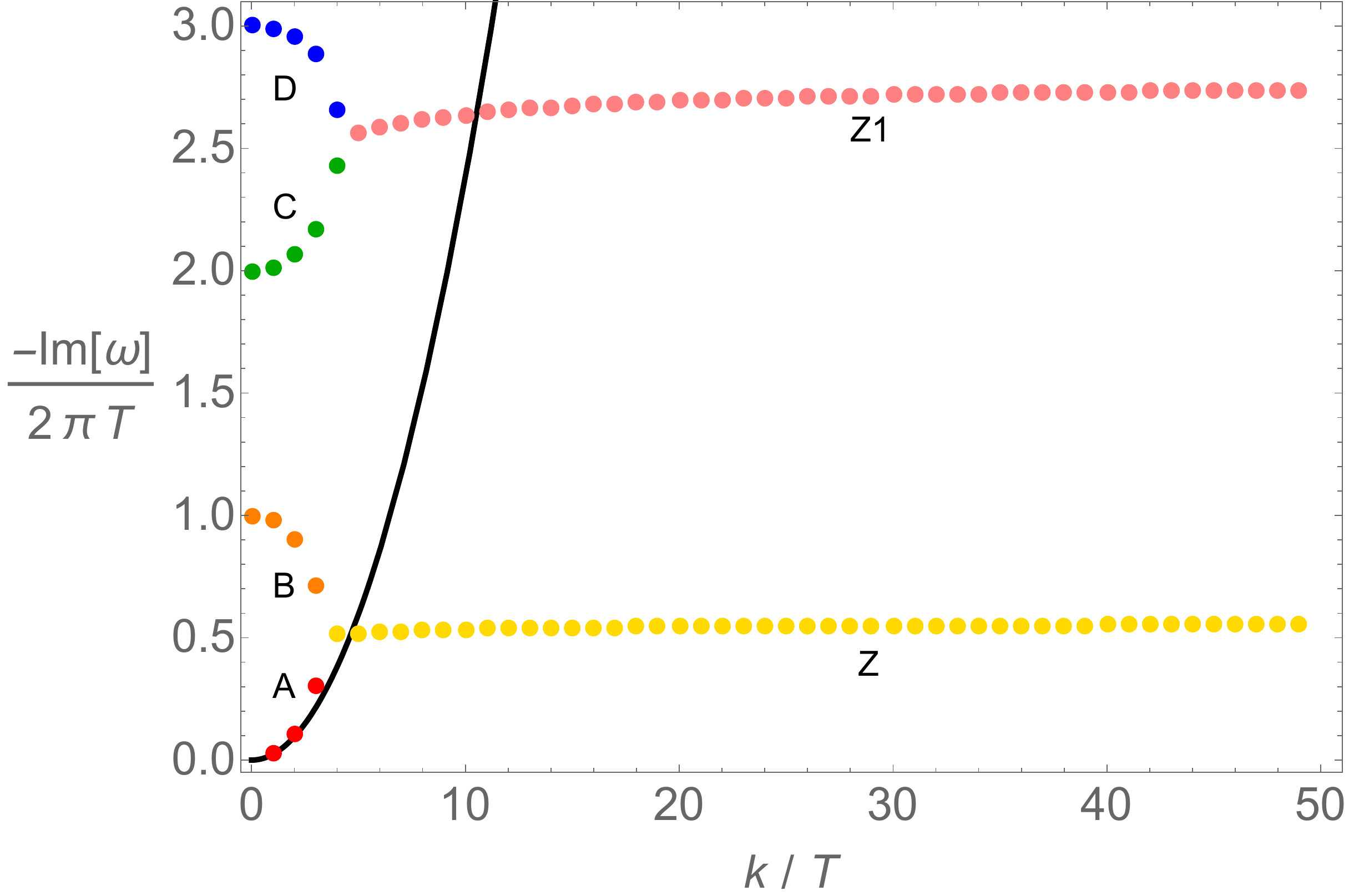} \label{FIGcol1}}
     \subfigure[$m/T = 10^2$ (intermediate $T$)]
     {\includegraphics[width=4.8cm]{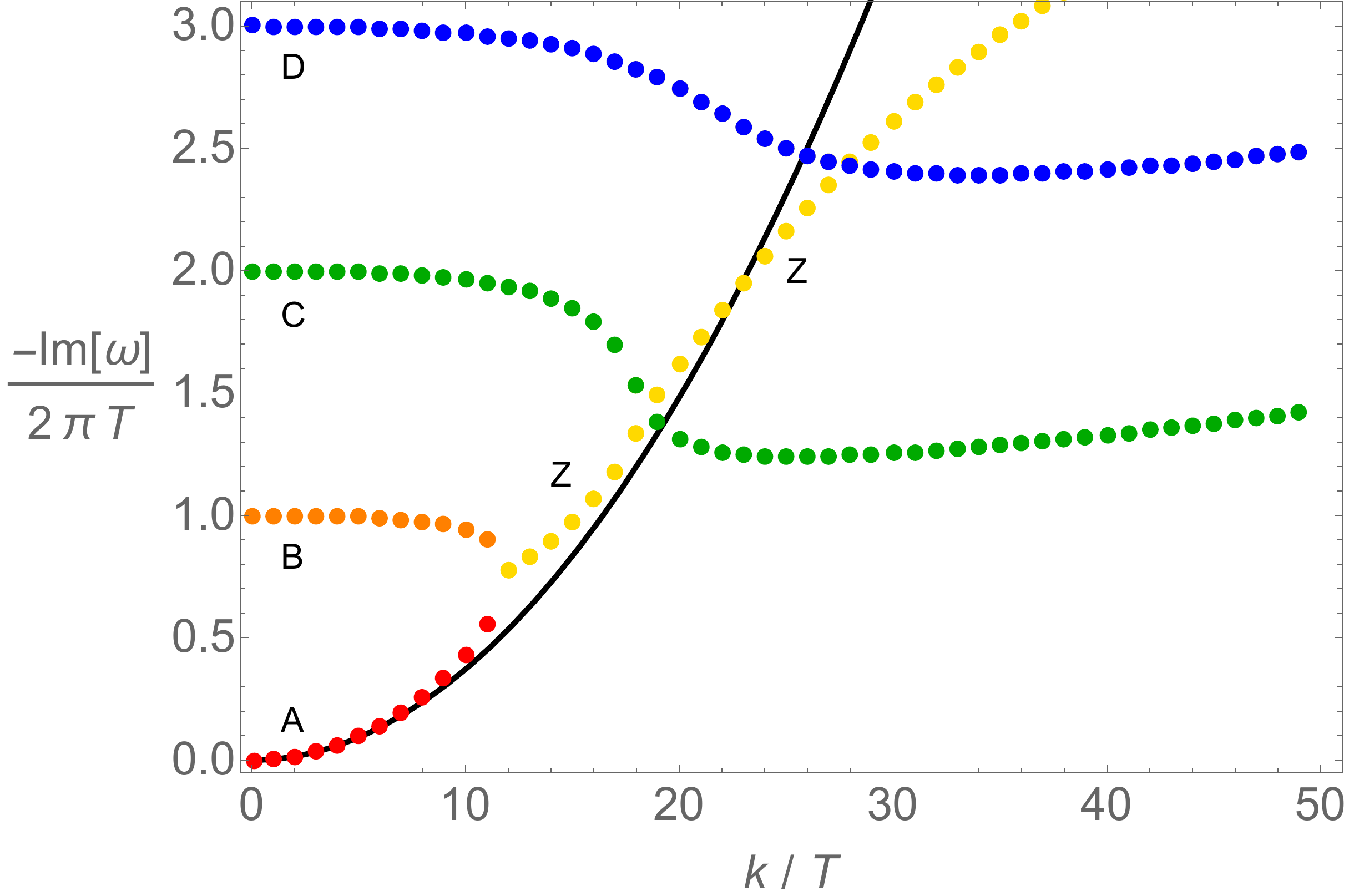} \label{FIGcol3}}
          \subfigure[$m/T = 10^3$ (low $T$)]
     {\includegraphics[width=4.8cm]{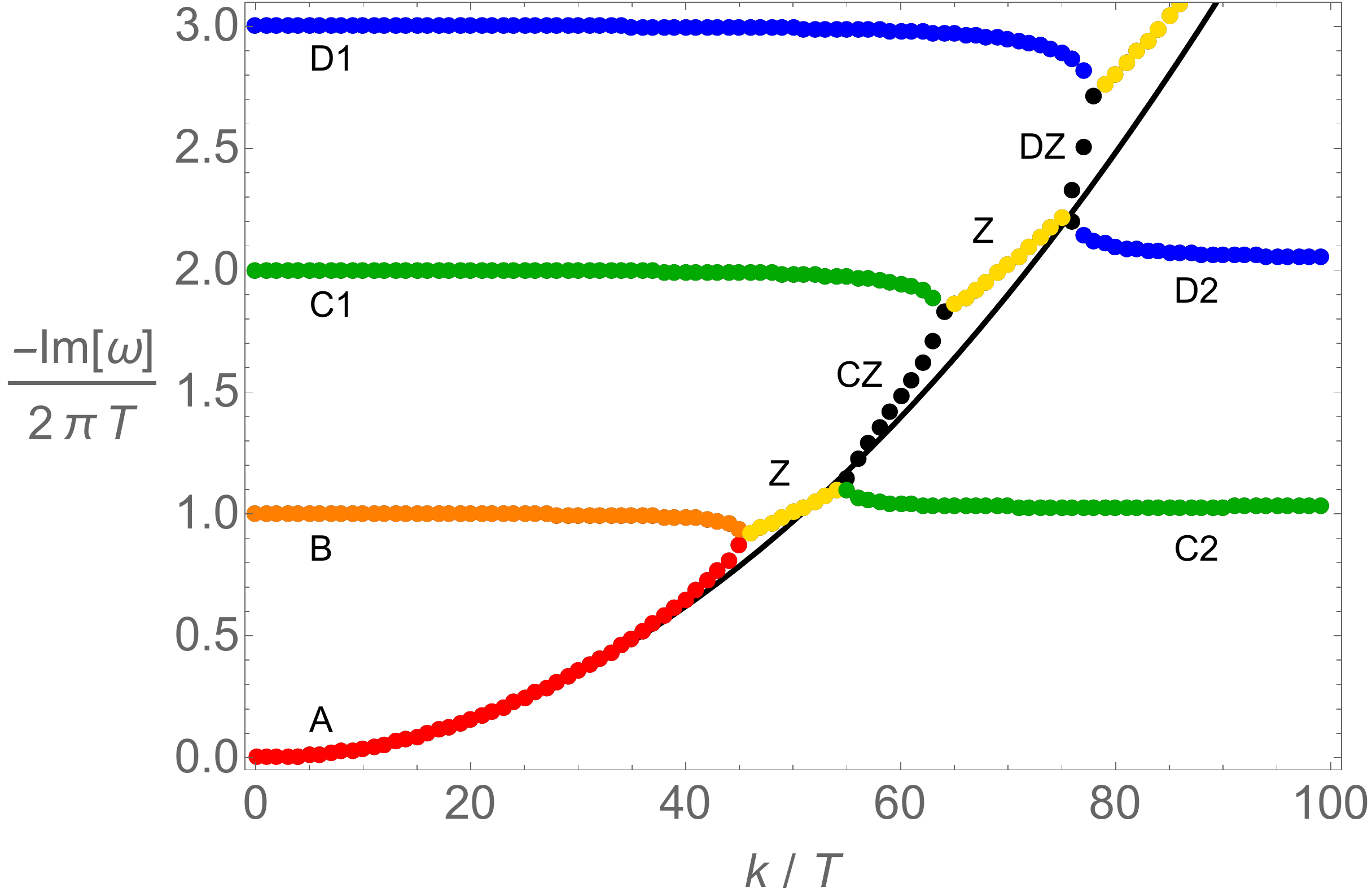} \label{FIGcol2}}
          \caption{Quasi-normal modes from high $T$ to low $T$. All dots are quasi-normal modes and the black solid line is the charge diffusion mode \eqref{eq2.11}. Note that the region around $k/T\sim50$ in Fig. \ref{FIGcol2} corresponds to Fig. \ref{FIG6a}.} \label{FIGcol}
\end{figure}

\paragraph{Quasi-normal modes at high $T$:} 
First, let us show the representative result of high $T$ in Fig. \ref{FIGcol1}.\footnote{At $m/T=0$, one can obtain the quasi-normal modes at higher temperature ($T\rightarrow\infty$). However, the qualitative structure of the quasi-normal modes does not change, i.e., Fig. \ref{FIGcol1} can be considered as the high enough temperature case.}
At high $T$, one can find that there is a feature in the quasi-normal modes: the structure consisting of three lines (e.g., $A$ (red), $B$ (orange), $Z$ (yellow)) repeats in the frequency $\omega$ direction.
As we decrease the temperature, the structure in Fig. \ref{FIGcol1} would change into what we observed in Sec.~\ref{sec3} (e.g., Fig. \ref{FIG2}). Now one may notice that it is non-trivial to obtain the low $T$ result (e.g., Fig. \ref{FIG2}) from the high $T$ result (e.g., Fig. \ref{FIGcol1}).

We find that as $T$ decreases, $Z$ (yellow) in Fig. \ref{FIGcol1} plays an important role in obtaining the low $T$ result. In particular, the low $T$ result can be understood by the interaction between $Z$ and the non-hydrodynamic mode ($C$ and $D$) in Fig. \ref{FIGcol1}.
We explain this further in two steps: i) $Z1$ (pink) becomes irrelevant; ii) $Z$ (yellow) interacts with non-hydrodynamic modes.

\begin{figure}[]
 \centering
     \subfigure[$m/T = 10$]
     {\includegraphics[width=4.83cm]{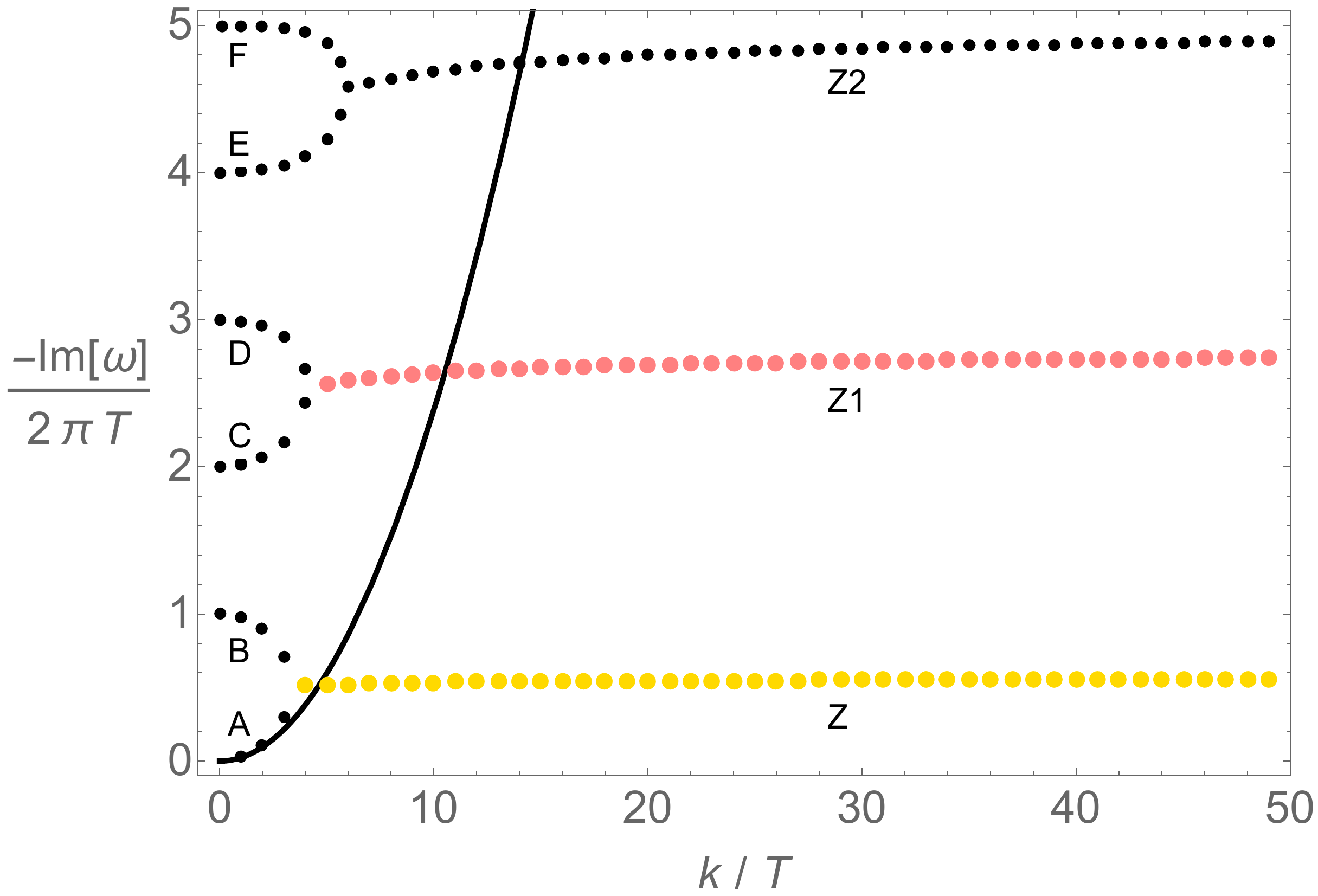} \label{FIG8a}}
     \subfigure[$m/T = 13$]
     {\includegraphics[width=4.83cm]{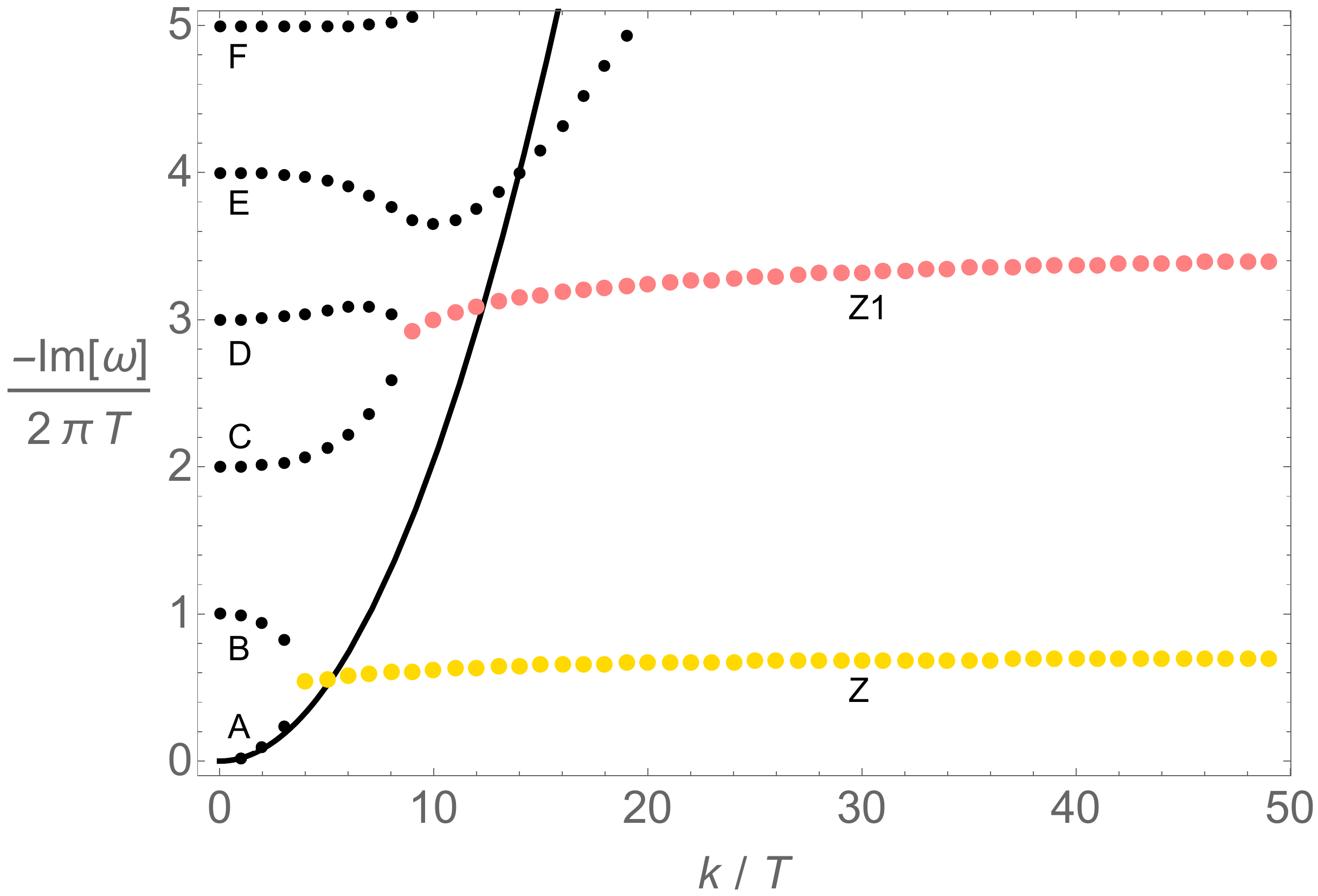} \label{FIG8b}}
          \subfigure[$m/T = 14$]
     {\includegraphics[width=4.83cm]{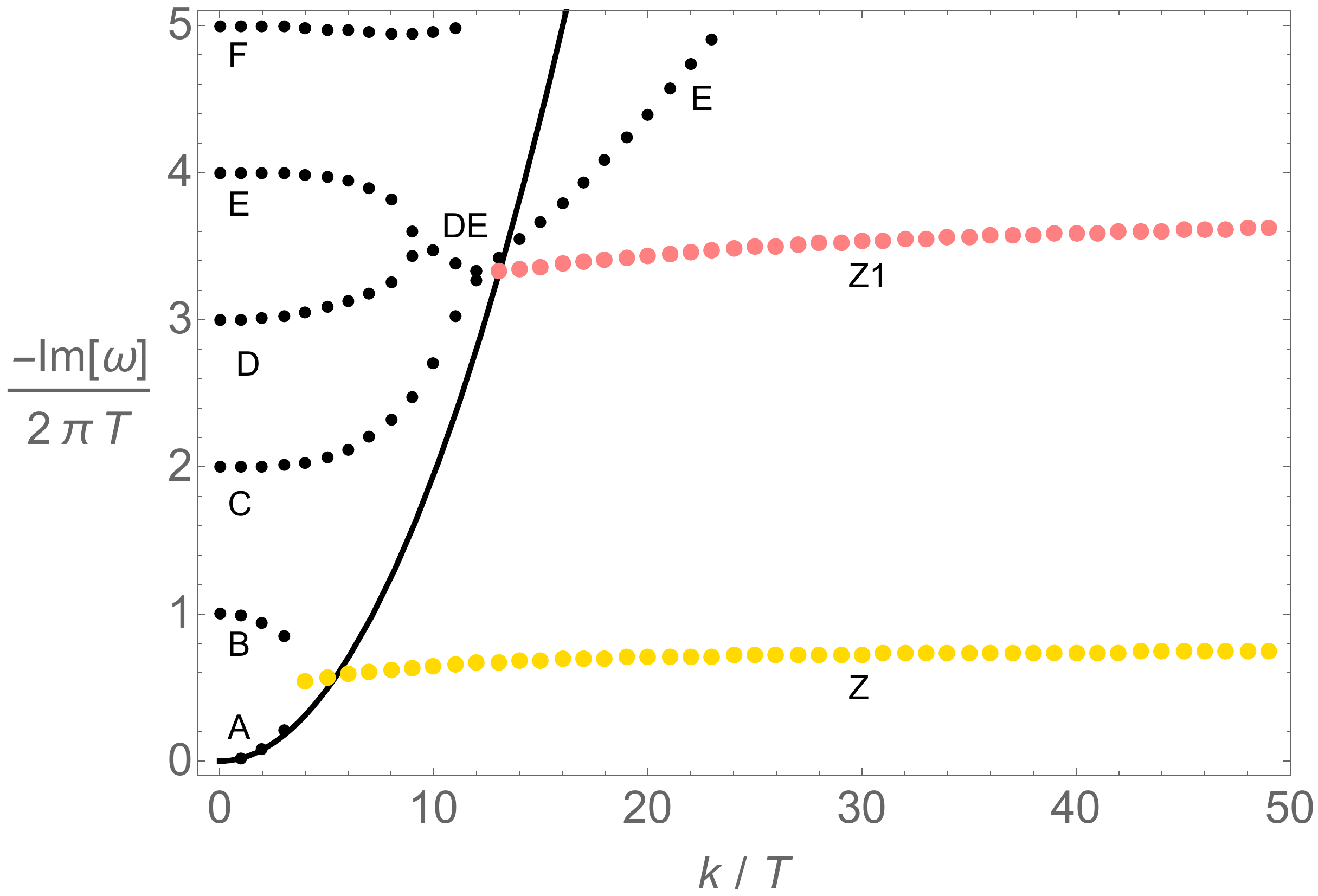} \label{FIG8c}}
          \subfigure[$m/T = 17$]
     {\includegraphics[width=4.83cm]{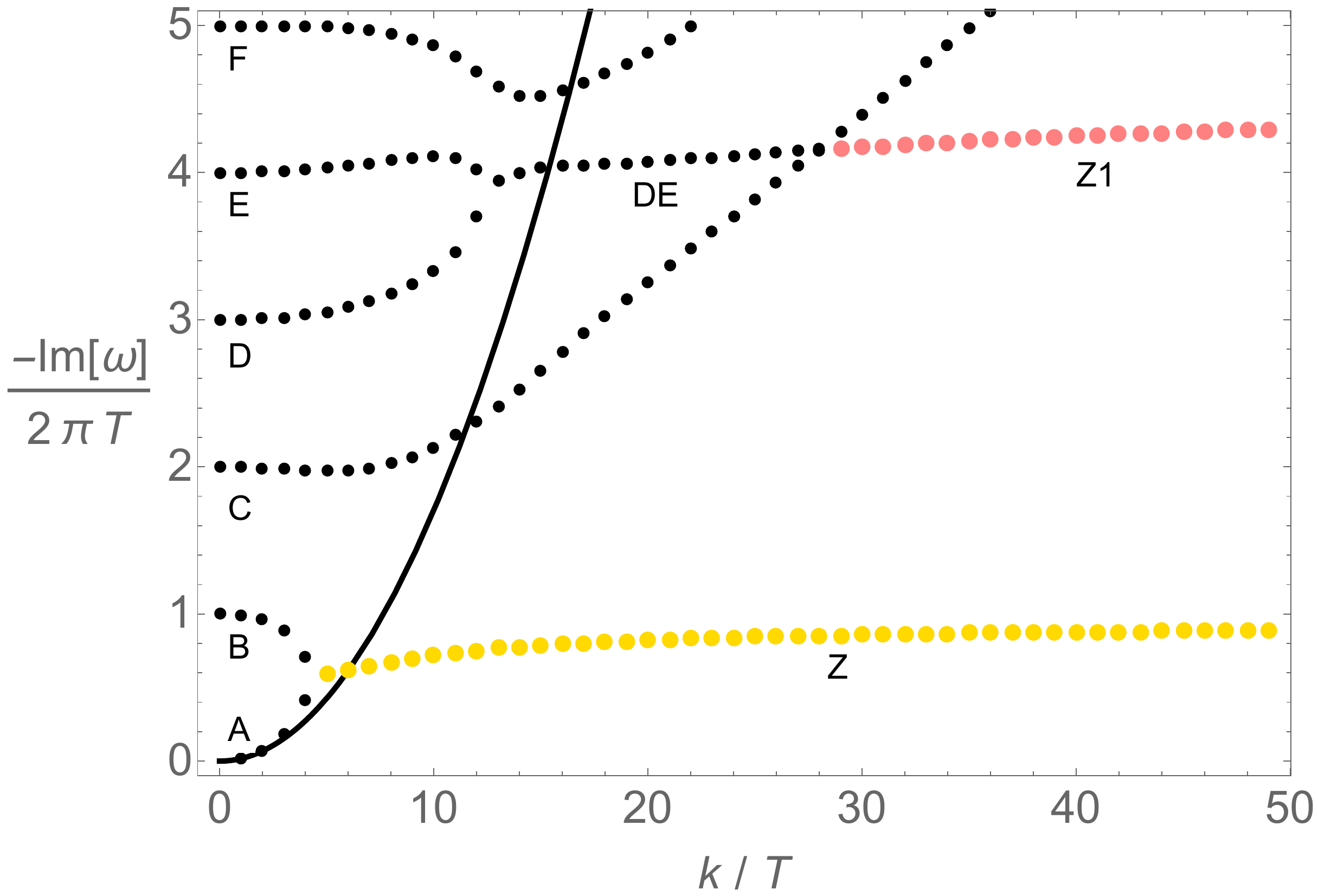} \label{FIG9a}}
     \subfigure[$m/T = 20$]
     {\includegraphics[width=4.83cm]{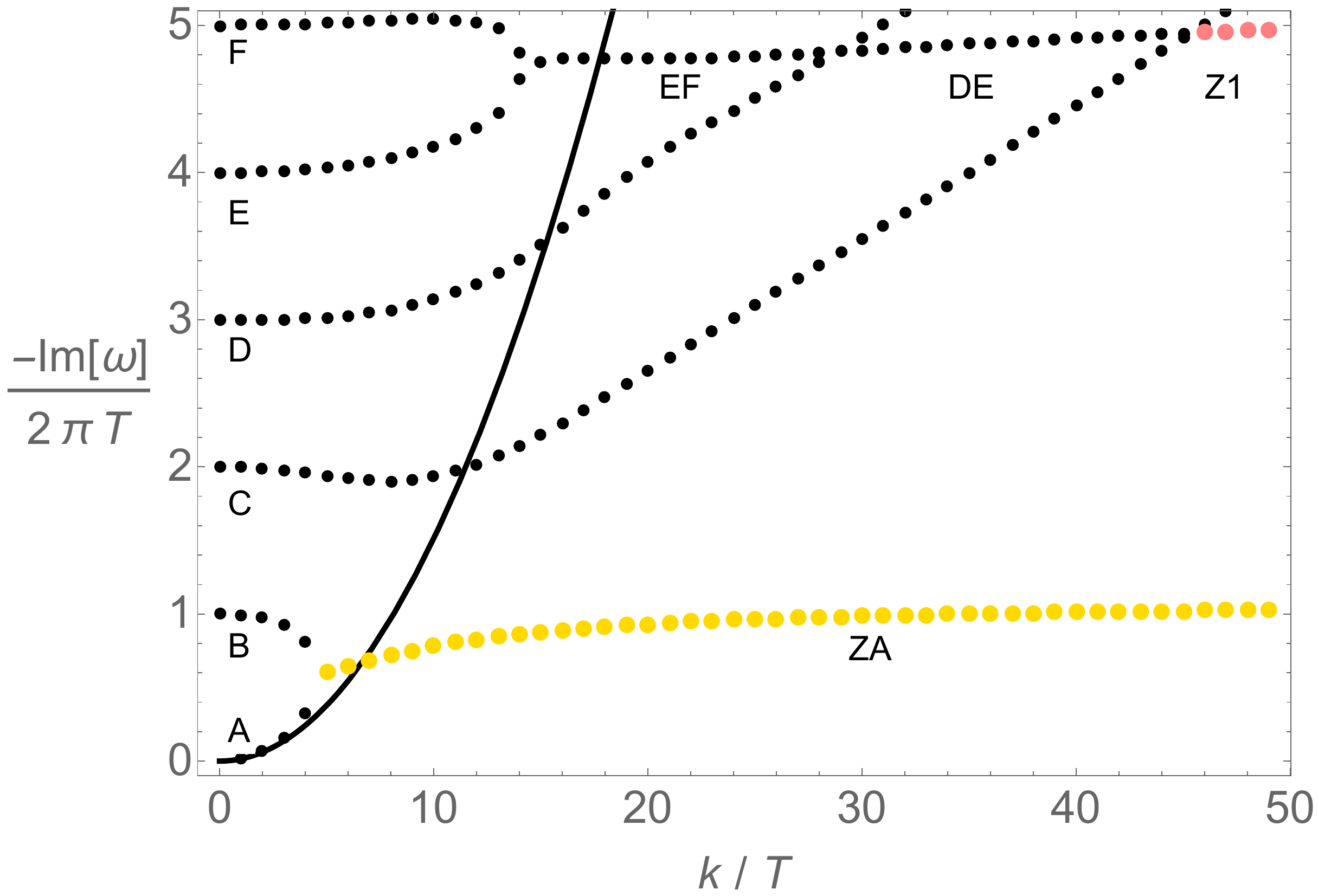} \label{FIG9b}}
          \subfigure[$m/T = 30$]
     {\includegraphics[width=4.83cm]{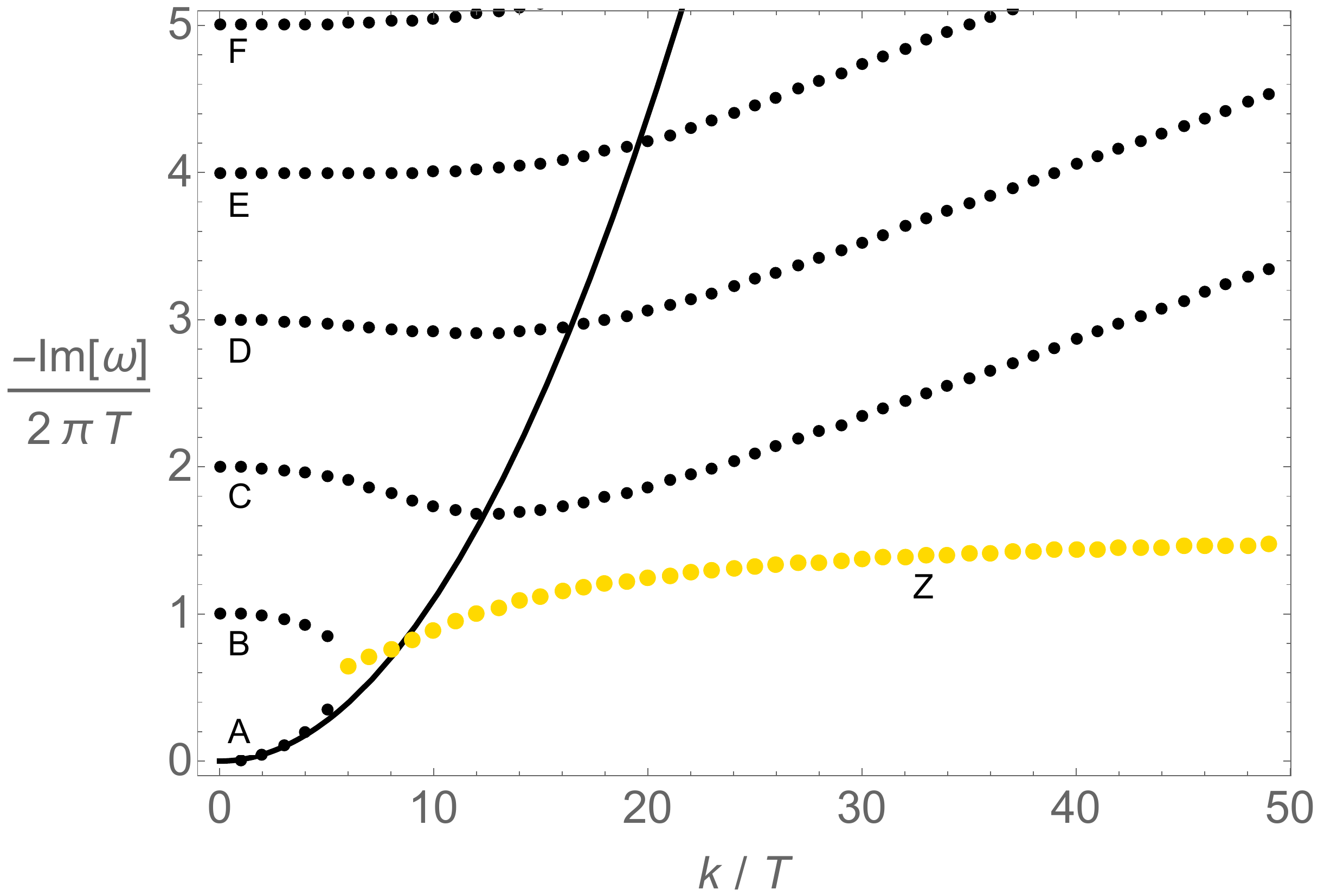} \label{FIG9c}}
          \caption{Quasi-normal modes from high $T$ to intermediate $T$. $DE$ in Fig. \ref{FIG8c} is made by the collision between $D$ and $E$. $EF$ in Fig. \ref{FIG9b} is made by the collision between $E$ and $F$.} \label{FIG8}
\end{figure}

\paragraph{From high $T$ to intermediate $T$:} The first step can be seen in the intermediate $T$ regime. In Fig. \ref{FIG8}, as $T$ is lowered from Fig. \ref{FIG8a} to Fig. \ref{FIG9c}, one can see that $Z1$ (pink) is interacting with the non-hydrodynamic mode  ($E$) and goes in the higher frequency regime (i.e., $Z1$(pink) is moving upward from Fig. \ref{FIG8a} to Fig. \ref{FIG9c}). Thus, at intermediate $T$ (e.g., Fig. \ref{FIG9c}), $Z1$ does not contribute to the low $T$ result.

\paragraph{From intermediate $T$ to low $T$:} 
Based on the previous paragraph, we have the representative result of intermediate $T$ in Fig. \ref{FIGcol3}. Comparing Fig. \ref{FIGcol3} and Fig. \ref{FIGcol2}, now one can see that $Z$ is important for the low $T$ result.
For instance, interacting with $C$ (green), $Z$ (yellow) results in two major consequences:
\begin{itemize}
\item{\textbf{Splitting effect}: $Z$ is splitting $C$ into two parts ($C1$, $C2$).}
\item{\textbf{Appearance of new line}: while splitting, $Z$ makes a new line $CZ$ (black dots).}
\end{itemize}
Similar behavior also occurs in the interaction between $D$ (blue) and $Z$ (yellow), i.e., $D$ is separated into two parts ($D1$, $D2$) and $DZ$ appears.

The splitting effect is important for the IR modes \eqref{eq3.10} (e.g., for the first non-hydrodynamic mode, one needs not only $B$, but also $C2$) and the appearance of new line has its significance for the hydrodynamic mode \eqref{eq2.11} (e.g., at lower $T$, $CZ$ and $DZ$ will be matched with \eqref{eq2.11}).
Therefore, $Z$ in high $T$ plays an important role for the low $T$ result.


\bibliographystyle{JHEP}

\providecommand{\href}[2]{#2}\begingroup\raggedright\endgroup

\end{document}